\documentclass[
superscriptaddress,
groupedaddress,
nofootinbib,
nobibnotes,
amsmath,
amssymb, 
aps, 
longbibliography,
10pt,
rmp, 
reprint, 
floatfix 
]{revtex4-2}

\usepackage{amsmath}

\usepackage{graphicx}
\usepackage[table]{xcolor}  
\usepackage{verbatim}

\usepackage[colorlinks,
            linkcolor={red!70!black},
            citecolor={blue!60!black},
            urlcolor={blue!80!black}]{hyperref}

\newcommand{\avg}[1]{\left\langle#1\right\rangle} 
\newcommand{\orderof}[1]{\ensuremath{\mathcal{O}\!\left[#1\right]}} 

\definecolor{mauricio}{RGB}{21,168,48}
\definecolor{supercritical}{RGB}{140,128,36}
\definecolor{subcritical}{RGB}{95,163,225}
\definecolor{mypink}{RGB}{255,0,225}

\definecolor{matheus}{RGB}{0,122,0}
\definecolor{denis}{RGB}{211,18,55}
\definecolor{constantino}{RGB}{55,18,255}
\def\hlinewd#1{%
\noalign{\ifnum0=`}\fi\hrule \@height #1 %
\futurelet\reserved@a\@xhline}

\usepackage{listings}
\newcounter{exampleproblem}
\renewcommand{\theexampleproblem}{\thesubsection.\arabic{exampleproblem}} 
\newcommand{\exampleproblem}{
  \refstepcounter{exampleproblem}
  \textbullet\space\textit{Problem \theexampleproblem.}\space
}

\NewDocumentCommand{\netstate}{o}{\ensuremath{\vec{\sigma}\IfNoValueTF{#1}{}{^{(#1)}}}}
\NewDocumentCommand{\memstate}{o}{\ensuremath{\vec{\xi}\IfNoValueTF{#1}{}{^{(#1)}}}}
\NewDocumentCommand{\spinnetstate}{o}{\ensuremath{\sigma\IfNoValueTF{#1}{}{^{(#1)}}}}
\NewDocumentCommand{\spinmemstate}{o}{\ensuremath{\xi\IfNoValueTF{#1}{}{^{(#1)}}}}
\NewDocumentCommand{\overlapmem}{o}{\ensuremath{M\IfNoValueTF{#1}{}{\!\!\left(#1\right)}}}

\NewDocumentCommand{\transpose}{o m}{%
  \IfNoValueTF{#1}
    {\ensuremath{#2^{\rm T}}}
    {\ensuremath{\left(#2\right)^{\rm T}}}
}

\newcommand{\ctalk}{\kappa_i^{(\nu)}}
\newcommand{\cctalk}{C_i^{(\nu)}}
\newcommand{\qEA}{q_{\rm EA}}
\newcommand{\mmax}{{\rm max}}
\newcommand{\Perr}{p_{\rm err}}
\newcommand{\Hamilt}{\mathcal{H}}
\newcommand{\sign}{{\rm sign}}
\newcommand{\erf}{{\rm erf}}
\NewDocumentCommand{\Hfunc}{o}{\ensuremath{\Hamilt\IfNoValueTF{#1}{}{\!\left(#1\right)}}}
\NewDocumentCommand{\Ffunc}{o}{\ensuremath{{\rm F}\IfNoValueTF{#1}{}{\!\left(#1\right)}}}
\NewDocumentCommand{\dHamm}{o}{\ensuremath{d_{\rm H}\IfNoValueTF{#1}{}{\!\left[#1\right]}}}

\usepackage{etoolbox}
\makeatletter
\patchcmd{\frontmatter@abstract@produce}
  {\vskip200\p@\@plus1fil
   \penalty-200\relax
   \vskip-200\p@\@plus-1fil}
  {}
  {}
  {}
\makeatother

\usepackage{silence}
\begin{document}
\setcitestyle{numbers,super,open={},close={}}

\title{Learning About Learning: A Path from Spin Glasses to Artificial Intelligence}

\author{Denis D. Caprioti} 
\thanks{These authors contributed equally to this work}
\affiliation{NeuroPhysics Lab, Departamento de Física, Universidade Federal de Santa Catarina, Florianópolis SC, Brazil 88040-900}

\author{Matheus Haas}
\thanks{These authors contributed equally to this work}
\affiliation{NeuroPhysics Lab, Departamento de Física, Universidade Federal de Santa Catarina, Florianópolis SC, Brazil 88040-900}

\author{Constantino F. Vasconcelos}
\affiliation{NeuroPhysics Lab, Departamento de Física, Universidade Federal de Santa Catarina, Florianópolis SC, Brazil 88040-900}

\author{Mauricio Girardi-Schappo}
\thanks{Corresponding author}
\email{girardi.s@gmail.com}
\affiliation{NeuroPhysics Lab, Departamento de Física, Universidade Federal de Santa Catarina, Florianópolis SC, Brazil 88040-900}

\date{\today}

\begin{abstract}
The Hopfield model, originally inspired by spin glasses, occupies a central place at the
intersection of statistical mechanics, neural networks, and  artificial intelligence.
Despite its conceptual simplicity and broad applicability, it is rarely integrated into the  undergraduate
physics curriculum. We present the Hopfield model as a pedagogically rich framework
that naturally unifies core topics from the undergraduate physics curriculum and  that provides a concise    introduction based on
  concepts such as a  model's energy function, dynamics, and pattern stability. We  discuss
 practical aspects of its simulation and provide simulation codes.
We also propose problems designed to mirror research practice, which can be
included in undergraduate classes.
\end{abstract}

\maketitle

\section{Introduction}

In the 1980s, John Hopfield~\cite{Hopfield1982,Hopfield1984} introduced a neural network model for associative memory.
Given an input, the network evolves toward an associated pattern encoded in the connections between neurons.
The model was inspired by the physics of spin glasses and remains
a central paradigm for memory encoding in the brain.\cite{RollsTreves1999Book,RollsTreves2024}
The Hopfield model can also be used to compute near-optimal solutions to complex combinatorial problems
by mapping a cost function onto the network Hamiltonian.\cite{HopfieldTank1985Opt1,HopfieldTank1986Opt2,hertzRedesNeurais,Rojas1996}
This work led to Hopfield sharing the 2024 Nobel Prize in Physics with Geoffrey Hinton,
whose work focused on Boltzmann machines.\cite{Nobel2024,HintonSej1983,AckleyHintonSej1985}

Neural networks have rapidly evolved from  academic tools to widely used technologies central
to data analysis, artificial intelligence (AI), and decision making~\cite{hertzRedesNeurais,Rojas1996,Nobel2024,TobochnikGould2026}
and has increased the demand for physicists in
both research and industry who can apply quantitative methods to data analysis and problem solving.
Although undergraduate physics provides many of the necessary conceptual foundations,
these topics are often taught in isolation and without clear connections to contemporary applications.
As a result, students may not  recognize their broader relevance.
The Hopfield model offers a compact and pedagogically rich framework
in which these connections can be made explicit.
Through this simple model, we discuss how core undergraduate concepts
from statistical physics, computational methods, dynamical systems, and linear algebra relate to AI.
By linking fundamental physics to contemporary AI applications,
our goal is to help students understand, apply, and critically
engage with the computational tools increasingly central to research, industry, and society.

\begin{figure*}[t!]
\centering
 \includegraphics[width=1\linewidth]{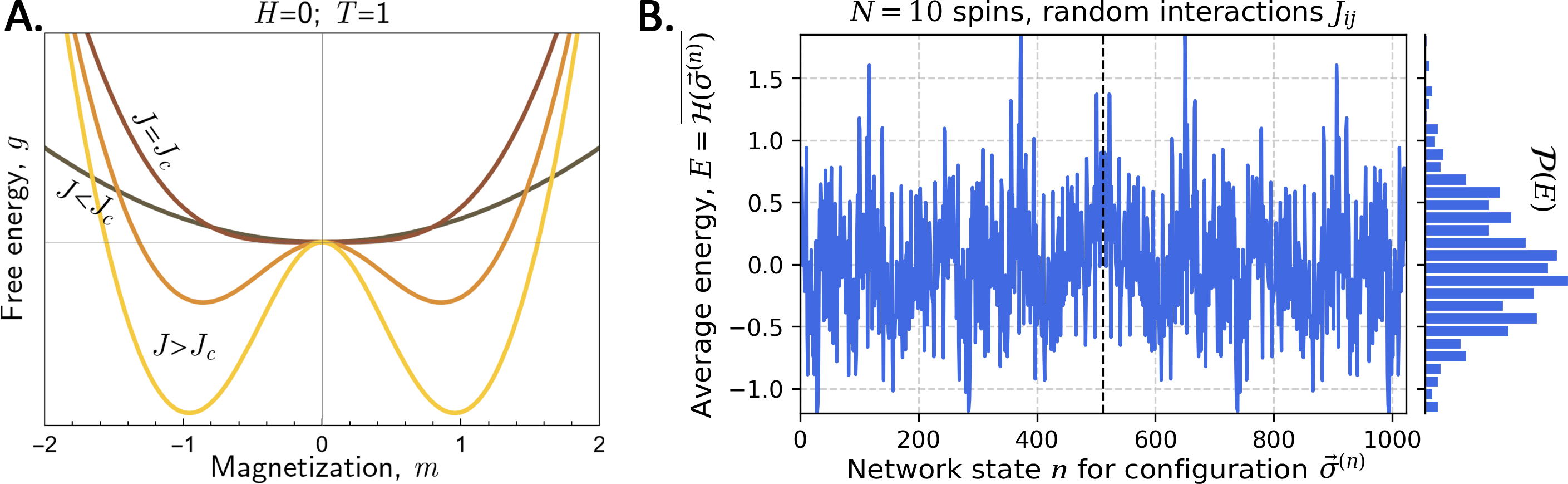}
 \caption{\label{fig:isingenergy}(a) The thermodynamic states corresponding to minima of the Gibbs free energy depend on $J$.
    More minima appear
    in the spin glass, where $J$ is no longer homogeneous.
    (b) The energy of a spin glass is shown
    for all $2^N=1024$ microscopic states of a system with $N=10$ spins, with $H_i=0$.
    Each local minimum is a metastable  state.
    The curve was obtained by averaging Eq.~\eqref{eq:SKHamilton} over 100 realizations
    of the interaction matrix, $\mathbf{J}$.
    $J_{ij}$ was sampled from a Gaussian distribution
    with zero mean and unit standard deviation.
    For display purposes, network configuration $\netstate[n]$ 
    is assigned an index $n$ and is plotted on the horizontal axis.     The energy landscape approximately follows a Gaussian distribution (right) with $\avg{E}\approx0$, which motivates
    random-energy models for spin glasses.~\cite{Derrida1980REM,Ruelle1987}
    The dashed line marks the state $n=2^9=512$. States labeled by $n>512$ correspond
    to spin-flipped configurations, producing the mirror symmetry observed in the plot.}
\end{figure*}

Because of its interdisciplinary nature, the Hopfield model is not
typically covered in standard undergraduate courses, although it does appear in textbooks aimed at a more specialized audience.\cite{hertzRedesNeurais,peretto,Rojas1996,GouldTobochnikBook2007}
Section~\ref{sec:statphys} reviews the statistical physics of the Ising model and spin
glasses. 
In Sec.~\ref{sec:hopfield}, we introduce the Hopfield model and use
concepts from spin-glass theory to motivate the construction of its energy function,
showing how equilibrium states can be shaped through the encoding of memory patterns,
i.e., by training the network.
The dynamics of the Hopfield model are presented in Sec.~\ref{sub:dynamics},
and the stability of  equilibria is analyzed in Sec.~\ref{sub:patternstability}.
Section~\ref{sub:simulation} discusses how to simulate the model and code is available in the supplemental material. 
In Sec.~\ref{sub:biology}, we discuss the biological background of the Hopfield model.
Finally, Sec.~\ref{sec:examples} presents examples of how the Hopfield model can
be incorporated into various undergraduate physics courses, along with a set of problems intended to guide instructors.

\section{Statistical Physics: Ising ferromagnet and spin glass}
\label{sec:statphys}

Spin glasses are magnetic systems in which the interactions between spins are random and frustrated,
preventing the system from settling into a single simple ordered state.\cite{EdwardsAnderson1975,SherringtonKirk1975}
Instead, the system becomes trapped in one of many possible configurations,
forming a complex energy landscape with numerous locally stable states.
During the 1970s and 1980s, physicists were mainly interested in deriving the thermodynamic
properties of these systems analytically.\cite{Nishimore2001spinGlass}
In practice, this interest meant understanding how many stable states exist and how robust they are,
rather than determining which microscopic configurations are
associated with a given macroscopic state.\cite{Mezard1987book} 
We will discuss this thermodynamic perspective in this section.

The Ising model has $N$ spins $\sigma_i=\pm1$ which form a configuration $\netstate=\transpose{[\sigma_1\ \cdots\ \sigma_N]}$.
The Sherrington-Kirkpatrick mean-field spin glass Hamiltonian is~\cite{SherringtonKirk1975}
\begin{equation}
\label{eq:SKHamilton}
\Hfunc[\netstate]=-\dfrac{1}{2}\sum_{i,j}\dfrac{J_{ij}}{N}\sigma_i\sigma_j-\sum_{i=1}^N H_i\sigma_i\ ,
\end{equation}
where every spin interacts with every other spin. 
Positive coupling, $J_{ij}>0$, favors the alignment of spins $i$ and $j$
by contributing a negative term to the interaction sum in Eq.~\eqref{eq:SKHamilton},
thereby lowering the energy.
A negative coupling, $J_{ij}<0$, favors spins pointing in opposite directions.

The standard Curie--Weiss Hamiltonian is recovered by setting $J_{ij}=J$,~\cite{Salinas2001Ingles}
yielding the usual mean-field model of a magnet.
Its thermodynamic properties can be derived from the canonical partition function (see the Supplementary Material).
At high temperatures, the system exhibits a single thermodynamic state -- the paramagnet, corresponding to $m=0$.
As the temperature decreases, this minimum splits into two,
corresponding to magnetized states, $m=\pm m_0$ [see  Fig.~S1A in the Supplementary Material].
This symmetry break occurs at the critical temperature, $T_c=1/J$, and zero field, $H=0$.

Because $J$ is a property of the material, it is usually treated as fixed, and its role can go unnoticed.
However, the energy minima  depend on $J$ [Fig.~\ref{fig:isingenergy}(a)].
An example of how the couplings can affect the macroscopic state of
a system is provided by spin glasses.
In these systems, the homogeneous interaction $J$
is replaced by a randomly distributed  $J_{ij}$ around the average $\bar{J}=0$.
Because each spin interacts with many others, the interactions become frustrated:
some couplings favor spin $i$ pointing up, while others favor it pointing down.
This ambiguity leads to a random energy landscape with multiple metastable states
(each being an energy minimum) -- see Fig.~\ref{fig:isingenergy}(b).

\begin{figure*}[t!]
\centering
    \includegraphics[width=1\linewidth]{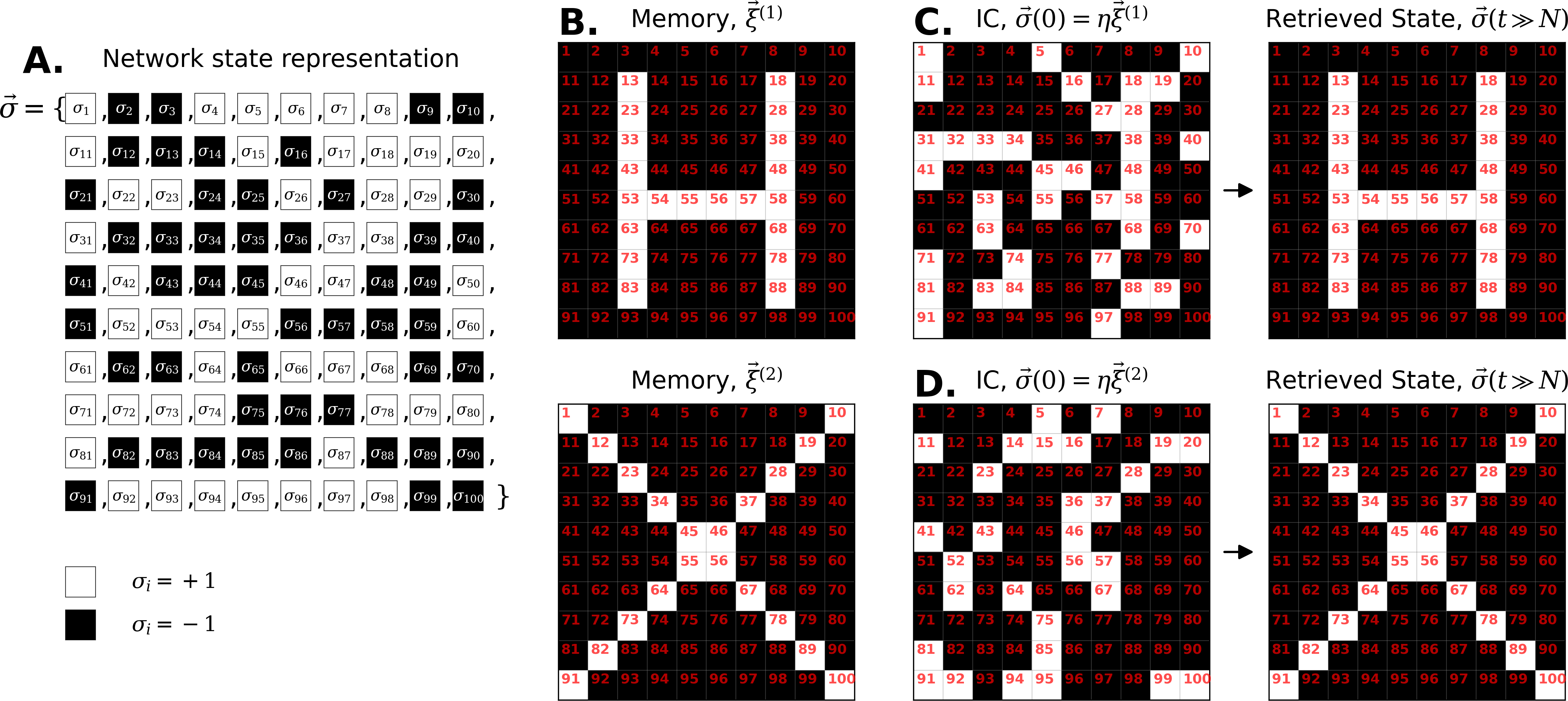}
    \caption{\label{fig:retrieval}
    (a) The state vector $\netstate$ of a Hopfield network with $N=100$ neurons  represented on a two-dimensional lattice
    of linear dimension  $L=\sqrt{N}=10$.
    (b)~Two memories are stored in the network using Eq.~\eqref{eq:weighthopfield}; the number at each site is the
    index of the corresponding neuron which forms either an ``H'' pattern, memory $\memstate[1]$,
    or an ``X'' pattern, memory $\memstate[2]$.
    (c)~A random initial condition near ``H'' (left) converges to ``H'' (right).
    (d)~A random initial condition near ``X'' (left) converges to ``X'' (right).
    }
\end{figure*}

In a regular ferromagnet, the  order parameter is the net magnetization $m$:
for low temperatures, $m\neq 0$, and thermal fluctuations drive $m\to 0$ at high temperatures.
Spin glasses behave differently: it has a ferromagnet state
at intermediary temperatures separating two phases of zero magnetization:
the paramagnet phase ($T>T_c$) and the glass phase ($T\to0$).
Even when the net magnetization vanishes for the glass phase at small temperatures,
the system remains frozen in a specific microscopic configuration.\cite{SherringtonKirk1975,Mezard1987book}
The similarity between two such frozen states, $\mu$ and $\nu$, is measured by the overlap
\begin{equation}
\label{eq:spinglassoverlap}
q^{(\mu,\nu)}=\dfrac{1}{N}\sum_{i=1}^{N}m_i^{(\mu)}m_i^{(\nu)},
\end{equation}
where $m_i^{(\mu)}=\avg{\sigma_i^{(\mu)}}$ is the local magnetization of spin $i$ in state $\mu$.
The overlap plays a central role in spin-glass theory,
and defines an order parameter for the glass state.
In the Hopfield model, we will use it to define the energy function,
and measure how closely the network microscopic state matches
a stored memory.\cite{hertzRedesNeurais,peretto}

Hopfield realized that an energy landscape with multiple minima could be turned into a useful feature.
Rather than asking which states appear spontaneously in a disordered material,
he asked whether $J_{ij}$ could be designed so that the system
would converge to a specific desired microscopic configuration.
In this way, pairwise interactions could be used to encode macroscopic information.

The central idea is that the coupling matrix $\mathbf{J}$ determines which microscopic configurations
have the lowest energy and are therefore stable.
The Hopfield model exploits this principle by constructing $J_{ij}$ so that selected patterns,
called \textit{memories}, become stable energy minima.
This idea became one of the foundations of what is now known as
\textit{unsupervised learning}.\footnote{\textit{Unsupervised learning} refers to methods that
    identify patterns in data without labeled examples. In contrast, \textit{supervised learning}
    adjusts parameters such as $J_{ij}$ using known input--output pairs, similarly to data fitting.\cite{TobochnikGould2026}
}
It also provides a simple description of how memories might be stored in the brain
and how neural systems can perform tasks such as pattern completion
and object recognition from incomplete information.\cite{Hopfield1982,RollsTreves1999Book,RollsTreves2024}

The Hopfield model is often introduced as a simplified neural network, but
the model is still fundamental in neuroscience and computer science
research.\cite{Rojas1996,GerstnerBook2014,RollsTreves1999Book,RollsTreves2024}
As we will see, it is  equivalent to an Ising spin glass
with appropriately chosen spin interactions.\cite{hertzRedesNeurais,peretto,Rojas1996}
The process of matching an input similar to a stored memory corresponds
to the system evolving through a high-dimensional phase space toward
a stable equilibrium state,  known in nonlinear dynamics as a fixed-point attractor.

We will also see that this convergence is guaranteed by the construction of the pairwise interactions from
simple elementary vector operations. For this reason, the model is a paradigmatic example 
of  an \textit{attractor network}.\cite{Rojas1996,GerstnerBook2014}
Thus, we can understand the Hopfield model by taking advantage of ideas that are typically introduced
in undergraduate courses in thermodynamics and statistical physics, computational physics, dynamical systems, and linear algebra.
Each of these subjects provides a complementary view of the same underlying phenomenon:
content-addressable memory.

\section{Hopfield neural network: content-addressable memory}
\label{sec:hopfield}

The fundamental question asked by Hopfield~\cite{Hopfield1982} is whether
\textit{content-addressable memory} can emerge from the collective interactions
of many simple units (neurons). The human brain provides a compelling
example of this capability: when presented with an incomplete or degraded image
of a familiar face, we can often recognize the person almost immediately.
This task is called \textit{pattern matching}.

At first glance, it is straightforward to imagine a computer program that
performs a similar task. Given a partial image as input, the program  searches
through a database and returns the most similar stored image according to some
chosen metric. Such an approach relies on an exhaustive search
through all the images in the database.
As the number of stored images increases, this
procedure rapidly becomes inefficient.

This strategy is also fragile. If the input image differs systematically from all stored examples
-- for instance, if the person is wearing a hat or has grown a beard --
the program may fail completely.
In contrast, content-addressable memory in biological systems is both efficient and robust:
such changes rarely prevent us from recognizing a person.
This ability to recover a memory from incomplete or distorted information
is known as \textit{error correction}.
How can such behavior emerge from simple interacting elements?

So far, we have identified two important features.
The interactions between spins determine the observable macroscopic state of the system.
Also, when these interactions change from pair to pair,
the system can exhibit a thermodynamic state with zero net magnetization
in which the spins remain frozen in a specific configuration.
Can such ``frozen states'' be deliberately encoded as the Hamiltonian minima?
If so, a stochastic dynamics satisfying detailed balance,
such as Metropolis~\cite{barkemaMC} or Glauber~\cite{tomeoliveira2015}   dynamics,
should drive the system toward them, enabling
the network to perform pattern matching.

Hence, rather than focusing on the minimization of the free energy,
our attention shifts to the structure of the Hamiltonian itself and to the states that minimize it.
Starting from an initial condition $\netstate(0)$
-- for example, an incomplete or noisy face image --
the dynamics naturally evolves toward the frozen state $\netstate(t)=\memstate[\mu]$ as $t\gg1$.
$\memstate[\mu]$ is one of the microscopic configurations $\netstate[n]$ that minimizes $\Hfunc$ (Fig.~\ref{fig:retrieval}).
We refer to these stable states $\memstate[\mu]$ as \textit{memories} or \textit{patterns},
where $\mu=1,\dots,P$.
Note that in the Hopfield network context, a ``microscopic configuration'' (thermodynamic terminology)
can also be called a ``network state'' (dynamics terminology).

\begin{figure}[t!]
\centering
    \includegraphics[width=1\linewidth]{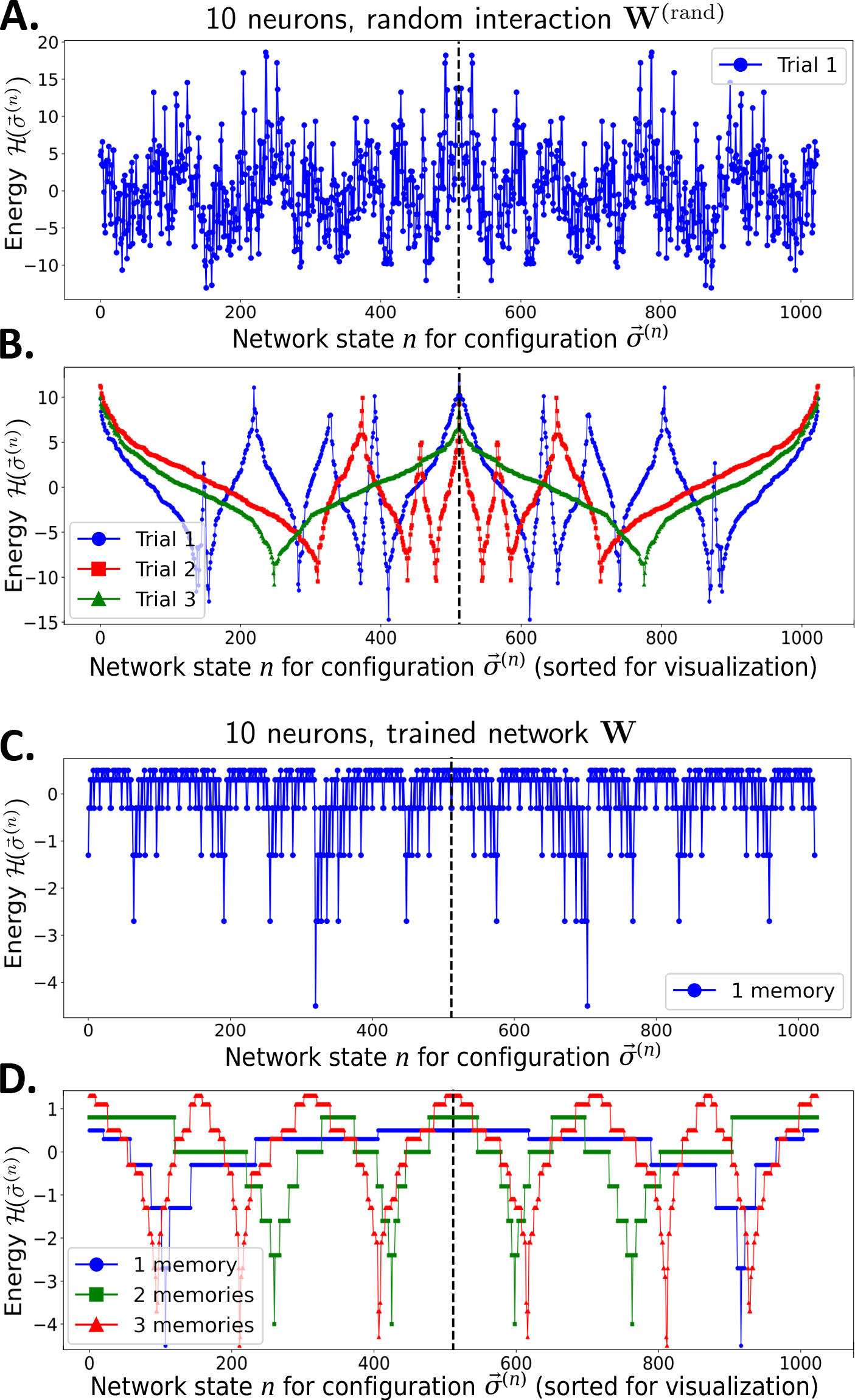}
    \caption{\label{fig:energylandscape}
    (a) The energy landscape defined by Eq.~\eqref{eq:energygeneral} with $H_i=0$,
     for all $2^N=1024$ states of a network with $N=10$ neurons.
    The weights $W_{ij}$ were randomly sampled from a  Gaussian distribution,
    producing a disordered landscape similar to that of a spin glass [Fig.~\ref{fig:isingenergy}(b)].
    (b) Three realizations of the random-interaction energy landscape.
    The network states $\netstate[n]$ were reordered according to energy similarity and arranged around the minima
    (decreasing/increasing energy to the left/right of each minimum).
    Each realization produces minima at different microscopic configurations,
    indicating different attractor states.
    (c) The energy landscape obtained after encoding a single memory (minimum) using Eq.~\eqref{eq:weighthopfield},
    with unsorted states along the horizontal axis.
    (d)~The energy landscapes for $P=1$, $P=2$, and $P=3$ encoded memories (minima),
    with states reordered to emphasize the minima.
    Because of the spin-flip symmetry of Eq.~\eqref{eq:energydef},
    each memory has a corresponding antimemory that is also an attractor.
    }
\end{figure}

\subsection{Constructing the energy function}
\label{sub:energy}

We consider a network with $N$ neurons and define $x_i=1$ when neuron $i$ is actively firing action potentials (spikes), corresponding to a nonzero firing rate.~\cite{GerstnerBook2014}
Otherwise, $x_i=0$ and the neuron is considered silent.
Biologically, silent neurons may still fire occasional random spikes.
This noisy activity is analogous to the effect of temperature in the ferromagnet,
and is discussed in Sec.~\ref{sub:biology}.
To connect with the Ising formalism, we define $\sigma_i=2x_i-1$, so that $\sigma_i=1$ represents an active neuron and $\sigma_i=-1$ an inactive one.
The microscopic network state is an $N$-dimensional vector
$\netstate = \transpose{\left[\sigma_1\ \cdots\ \sigma_N\right]}$,
analogous to the Ising model state, and  can be plotted as a
square lattice -- see Fig.~\ref{fig:retrieval}(a) --
even though the model is not two-dimensional.

A memory $\memstate[\mu]$ consists of a subset of the $N$ neurons firing together,
while the remaining neurons remain silent [Fig.~\ref{fig:retrieval}(b)].
The energy function derived in the following should not be interpreted as a physical energy
in the same sense as the Hamiltonian of a magnetic system.
Instead, it is a mathematical construct used to describe the collective dynamics of the network.
In the nonlinear dynamics literature, such a function is known as a Lyapunov function.~\cite{hertzRedesNeurais,peretto}

We define the overlap between a network state $n$ and a memory $\mu$ as,
\begin{align}
\label{eq:overlap}
\overlapmem[\memstate[\mu],\netstate[n]]&=\dfrac{1}{N}\memstate[\mu]\cdot\netstate[n]
=\dfrac{1}{N}\transpose[]{\memstate[\mu]}\netstate[n]\ \\
&=\dfrac{1}{N}\sum_{i=1}^N\spinmemstate[\mu]_i\spinnetstate[n]_i\ .
\label{eq:overlapb}
\end{align}
Equation~\eqref{eq:overlap} directly follows from Eq.~\eqref{eq:spinglassoverlap},
and is simply the normalized scalar product between $\netstate[n]$ and $\memstate[\mu]$ [Fig.~S3(a) in the supplementary material].
The normalization factor $1/N$ ensures that
$\overlapmem[\memstate[\mu],\netstate[n]]\in[-1,1]$.
When $\overlapmem[\memstate[\mu],\netstate[n]] = 1$,
the network state matches the memory exactly, $\netstate[n]=\memstate[\mu]$.
When $\overlapmem[\memstate[\mu],\netstate[n]] = -1$,
the network matches the corresponding antimemory, $\netstate[n]=-\memstate[\mu]$.
This antimemory arises from the spin-flip symmetry of the system, which makes only half of the $2^N$ states unique.
We can use these results to define the energy function.

Mean-field theory predicts
that the internal energy of a ferromagnet is~\cite{Salinas2001Ingles} $U=\avg{\Hfunc}=-\frac{1}{2}JNm^2$,
where we used $\langle\sum_{i,j}\sigma_i\sigma_j\rangle\approx N^2m^2$.
Because $|m|\leq1$, $U$ has minima at $m=\pm1$ as expected for a zero-temperature ferromagnet. 
For the Hopfield Model, we want a particular microscopic configuration $\memstate[\mu]$ to minimize the energy.
So, the minima must be located at $M=\pm1$ instead, i.e.,
at microscopic configurations $\netstate[n]=\memstate[\mu]$, suggesting that $\Hfunc\sim-\frac{1}{2}N\overlapmem^2$.
Hence, we define the energy function as~\cite{hertzRedesNeurais,peretto}
\begin{equation}
\label{eq:energydef}
\Hfunc = -\dfrac{1}{2}N \sum_{\mu=1}^P \left(\overlapmem[\memstate[\mu],\netstate]\right)^2\ .
\end{equation}
This energy is shown in Fig.~S3(b) for $P=1$.

We substitute Eq.~\eqref{eq:overlapb} into Eq.~\eqref{eq:energydef} to define the interaction weight (details in the supplementary material)
\begin{equation}
\label{eq:weighthopfield}
W_{ij} = \frac{1}{N} \sum_{\mu=1}^P \spinmemstate[\mu]_i \spinmemstate[\mu]_j\ .
\end{equation}
taking $W_{ii}=0$ with no loss of generality.~\cite{hertzRedesNeurais,peretto} Then Eq.~\eqref{eq:energydef} becomes
\begin{equation}
\label{eq:energygeneral}
\Hfunc[\netstate] = -\frac{1}{2} \sum_{i=1}^N \sum_{j=1}^N W_{ij} \sigma_i \sigma_j - \sum_{i=1}^N H_i \sigma_i\ ,
\end{equation}
where $H_i$ is a local field, which is biologically interpreted as the activation threshold of neuron $i$.
Note that making $J_{ij}=N\, W_{ij}$ turns the Hopfield energy function into the spin glass Hamiltonian, Eq.~\eqref{eq:SKHamilton}.

Calculating the weights in Eq.~\eqref{eq:weighthopfield} from a set of patterns
$\mathbb{P}=\left\{\memstate[\mu] : \mu=1,\cdots,P\right\}$
is called \textit{training} the network.
The weight $W_{ij}$ measures the correlation between the activity of neurons $i$ and $j$
across all stored patterns.
Equation~\eqref{eq:weighthopfield} therefore implements a generalized Hebbian learning rule:\cite{hebb1942}
connections are strengthened between neurons that tend to be active or silent  in the same memory $\mu$,
whereas other connections are weakened.
Mathematically, $W_{ij}$ increases when $\spinmemstate[\mu]_i=\spinmemstate[\mu]_j=\pm1$,
and decreases otherwise.

This mechanism is a simplified form of synaptic plasticity loosely inspired by biological learning processes.~\cite{deSchutterBook}
Positive couplings, $W_{ij}>0$, favor neurons sharing the same activity state,
and negative couplings, $W_{ij}<0$, favor opposite activity states and can be interpreted as effective inhibition between neurons.
Together, these interactions shape the energy landscape of the network,
allowing memories to be encoded and later retrieved (Fig.~\ref{fig:energylandscape}).
Other forms of encoding that work better for correlated memories exist,
and the reader is directed to Refs.~\citenum{hertzRedesNeurais,peretto,Sahoo2020} to learn more.

The Hopfield model can also be used to find near-optimal solutions to optimization problems.
To do so, the problem must be expressed as a cost function that can be mapped onto Eq.~\eqref{eq:energygeneral},
so that the solutions correspond to energy minima.
Terms proportional to $\sigma_i\sigma_j$ define the weights $W_{ij}$,
terms proportional to $\sigma_i$ define the biases $H_i$,
and constant terms do not affect the dynamics.
Cost functions containing higher-order terms involving three or more spins generally fall outside this framework.
Applications include graph bipartitioning, the traveling salesman problem,
the eight queens problem, weighted matching, and optimization tasks in image processing and pattern recognition.
Some of these examples are discussed in Refs.~\citenum{hertzRedesNeurais,Rojas1996}.

\subsection{Hopfield dynamics}
\label{sub:dynamics}
For the model to perform memory retrieval or solve optimization problems,
the network state $\netstate$ must evolve toward a minimum of $\Hfunc$.
Starting from an initial configuration $\netstate(0)$,
the dynamics should converge to a stored pattern $\netstate(t\gg1)=\memstate[\mu]$
after a number of iterations $t$.
This process is known as \textit{error correction}:
noise or distortions in the initial state are gradually removed,
and the network converges to the most similar memory.
In the example of face recognition, features such as a hat act as noise,
and the network can still recover the correct face.

Although the following dynamics can be derived more formally,~\cite{peretto}
we will introduce it phenomenologically, and then justify it by showing that it minimizes the energy function.
We want to find a relation between $\netstate(t)$ and its next value, $\netstate(t+1)$ assuming that the network evolves in discrete time steps.
We include the time dependence and write Eq.~\eqref{eq:energygeneral} as
\begin{equation}
\label{eq:energytime}
\Hfunc[t] = -\frac{1}{2}\sum_{i=1}^N h_i(t) \sigma_i(t)\ ,
\end{equation}
where
\begin{equation}
\label{eq:localfield}
h_i(t)=\sum_{j=1}^N W_{ij}\sigma_j(t) + H_i
\end{equation}
is the effective field acting on neuron $i$. We absorbed the prefactor $2$ into $H_i$,
because it is an arbitrary constant.

Given our intuition from a ferromagnet, we assume that
the state $\sigma_i$ of neuron $i$ needs to \textit{align} with the effective field $h_i(t)$
to minimize the energy. Hence, its next state must be
\begin{equation}
\label{eq:hopfielddynamics}
\sigma_i(t+1)\equiv\Ffunc[h_i(t)]=\Ffunc[\sum_{j=1}^{N}W_{ij}\sigma_j(t) + H_i]\ ,
\end{equation}
where $\Ffunc[x]=\sign(x)$ is   
defined by $\Ffunc[x\geq0]=+1$; $\Ffunc[x<0]=-1$.

By using the update rule in Eq.~\eqref{eq:hopfielddynamics},
we can compute the change in the energy function,
$\Delta\Hfunc[t]=\Hfunc[t+1]-\Hfunc[t]$,
produced by updating a single neuron $k$ chosen at random.
We use Eq.~\eqref{eq:energytime} to find
\begin{align}
\Delta\Hfunc[t] &= -\frac{1}{2}\left[\sum_{i=1}^N h_i(t+1) \sigma_i(t+1)
                                    -\sum_{i=1}^N h_i(t) \sigma_i(t)\right]\\
&= -\frac{1}{2}\left[h_k(t) \Delta\sigma_k(t) + \sum_{\substack{i=1\\i\neq k}}^N h_i(t) \Delta\sigma_i(t)\right]\\
&= -\frac{1}{2}h_k(t) \Delta\sigma_k(t)\ .
\label{eq:energyshift}
\end{align}
where $\Delta\sigma_i(t)=\sigma_i(t+1)-\sigma_i(t)$, and we used
the fact that we are updating only neuron $k$, so $\Delta\sigma_i(t)\equiv0$ for $i\neq k$, and $h_i(t+1)=h_i(t)$, because $W_{ii}=0$ (including for $i=k$).

When neuron $k$ changes its state,
the update rule in Eq.~\eqref{eq:hopfielddynamics} gives two distinct possibilities.
When $h_k(t)>0$, $\Delta\sigma_k(t)= 1-\sigma_k(t)=+2$. 
Likewise, when $h_k(t)<0$, we obtain $\Delta\sigma_k(t)= -1 - \sigma_k(t) = -2$.
In both cases, updating neuron $k$ decreases the energy, so $\Delta\Hfunc[t]<0$.
If the neuron state does not change, $\Delta\sigma_k(t)=0$ and therefore $\Delta\Hfunc[t]=0$.
Because neurons are updated one at a time, the network typically requires $t\sim\orderof{N}$ updates
to reach a state in which all neurons satisfy $\Delta\sigma_k(t)=0$, and
the dynamics has converged to an attractor corresponding to a minimum of $\Hfunc$.
For convenience, we define one Monte Carlo (MC) step as $N$ single-neuron updates;~\cite{barkemaMC}
in the neural network literature, $N$ updates is often called an ``epoch.''~\cite{Rojas1996}

So far, we have seen that the Hebb rule, Eq.~\eqref{eq:weighthopfield},
makes the intended memory states minima of the energy function, Eq.~\eqref{eq:energygeneral}.
We have also shown that the update rule, Eq.~\eqref{eq:hopfielddynamics},
drives the network toward minima of $\Hfunc$.
We still need to verify that a retrieved memory remains stable under further updates.
In other words, if the network reaches $\netstate(t)=\memstate[\mu]$,
then applying Eq.~\eqref{eq:hopfielddynamics} again should leave the state unchanged,
$\netstate(t+1)=\memstate[\mu]$.
If this condition holds, the memory $\memstate[\mu]$ is a stable attracting fixed point of the dynamics,
$\netstate(t+1)=\netstate(t)=\netstate^*=\memstate[\mu]$.

Note that training the network with Eq.~\eqref{eq:weighthopfield}
can unintentionally create local minima in $\Hfunc$
that do not correspond to any memory $\memstate[\mu]$
in the set $\mathbb{P}$.
This problem usually occurs when the number of stored memories $P$ becomes too large.
These unintended attractors are called \textit{spurious memories} or \textit{hallucinations}.
Although they are stable states, they make the network unreliable.
In practice, this behavior is analogous to recognizing the wrong person when shown a familiar face.
As we will see, this problem can be mitigated by limiting the number of stored memories.

\begin{figure*}[t!]
\centering
    \includegraphics[width=1\linewidth]{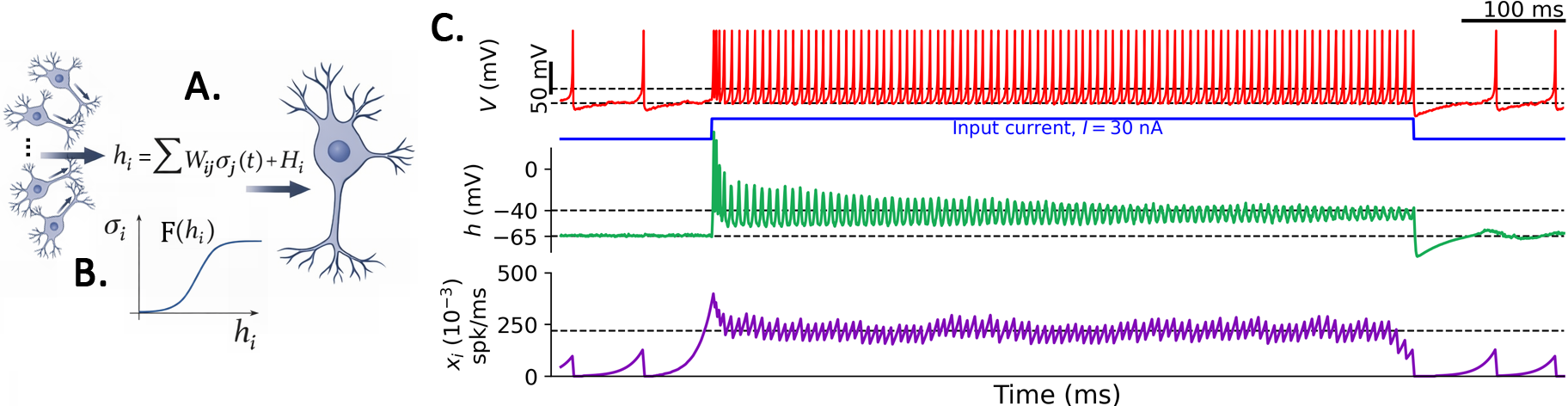}
    \caption{\label{fig:biology}
(a) Biological interpretation of Eq.~\eqref{eq:hopfielddynamics}.
    Post-synaptic potentials from other neurons are integrated by the membrane,
    yielding the effective field $h_i(t)$, where $H_i$ captures the activation threshold.~\cite{abbottNeuro,GerstnerBook2014,Shimoura2021Conn}
(b) The integrated signal is converted into the output $\sigma_i(t)=\Ffunc[h_i(t)]$,
    the corresponding firing rate is $x_i=(\sigma_i+1)/2$.
(c,~top) The membrane potential $V$ showing increased spiking frequency under a step input current.
    The dashed lines represent the resting voltage ($-65$\,mV) and spiking threshold ($\sim-40$\,mV).
(c,~middle) The corresponding effective field $h_i$ of the neuron in the top panel.
    It shifts from rest toward the spiking threshold upon stimulation,
    tracking all subthreshold membrane dynamics. It is just the membrane voltage $V$
    with spikes removed.
(c,~bottom) The firing rate $x_i$ estimated by convolving the spike train with an exponential kernel
    $K(t)=\Theta(-t)e^{t/W}/W$ ($W=60$~ms; $\Theta(t)\to$ Heaviside function),
    rising from $0$ to $\approx\!220$~spikes/ms (dashed line) during stimulation.
    The Hopfield model reduces the shown spiking activity to this all-or-none rate-coded output.
    }
\end{figure*}

\subsection{Pattern stability and storage capacity}
\label{sub:patternstability}

For the model to be useful, iterating Eq.~\eqref{eq:hopfielddynamics}
must drive the network toward one of the stored memories, $\memstate[\mu]$.
For $P=1$, it is straightforward to show that the stored pattern $\memstate$
is an attractor.
We use $W_{ij}=\xi_i\xi_j/N$ and set $\netstate(t)=\memstate$,
to obtain
\begin{align}
\sigma_i(t+1)&=\Ffunc[\sum_{j=1}^N\left(\frac{1}{N}\xi_i\xi_j\right)\xi_j + H_i]\\
&=\Ffunc[\xi_i\frac{1}{N}\sum_{j=1}^N\left(\xi_j\right)^2 + H_i]\\
&=\Ffunc[\xi_i + H_i]=\xi_i\ ,\label{eq:attractor1mem}
\end{align}
where we used $\xi_j^2=1$ and assumed $H_i\in(-1,1)$.
Under this condition, the sign of $\xi_i+H_i$ is always equal to the sign of $\xi_i$,
so the state remains unchanged after the update.
The same argument applies to $-\memstate$.

The situation becomes more subtle when multiple memories are stored.
Suppose we want to test whether a particular memory $\nu$ in the set $\mathbb{P}$ remains stable:
\begin{subequations}
\begin{align}
\sigma_i(t+1)&=\Ffunc[\sum_{j=1}^N\left(\frac{1}{N}\sum_{\mu=1}^P\spinmemstate[\mu]_i\spinmemstate[\mu]_j\right)\spinmemstate[\nu]_j + H_i]\\
&=\Ffunc[\frac{1}{N}\sum_{\mu=1}^P\spinmemstate[\mu]_i\left(\sum_{j=1}^N\spinmemstate[\mu]_j\spinmemstate[\nu]_j\right) + H_i]\,.
\end{align}
\end{subequations}
As before, the term with $\mu=\nu$ simplifies because $(\spinmemstate[\nu]_j)^2=1$.
We separate this contribution from the remaining patterns to find
\begin{multline}
\sigma_i(t+1)={\rm F}\!\Bigg(
\spinmemstate[\nu]_i\frac{1}{N}\sum_{j=1}^N\left(\spinmemstate[\nu]_j\right)^2\\
    +\frac{1}{N}\sum_{\substack{\mu=1\\\mu\neq\nu}}^P\spinmemstate[\mu]_i\bigg(\sum_{j=1}^N\spinmemstate[\mu]_j\spinmemstate[\nu]_j\bigg) + H_i \Bigg)\ ,
\end{multline}
which becomes
\begin{equation}
\label{eq:itercrosstalk}
\sigma_i(t+1)=\Ffunc[\spinmemstate[\nu]_i + \ctalk + H_i]\ ,
\end{equation}
where
\begin{equation}
\label{eq:crosstalk}
\ctalk=\sum_{\substack{\mu=1\\\mu\neq\nu}}^P\spinmemstate[\mu]_i
\left[\frac{1}{N}\sum_{j=1}^N\spinmemstate[\mu]_j\spinmemstate[\nu]_j\right]\ 
\end{equation}
is the \textit{crosstalk} term arising from interference between different stored memories.

\begin{figure*}[t!]
\centering
    \includegraphics[width=1\linewidth]{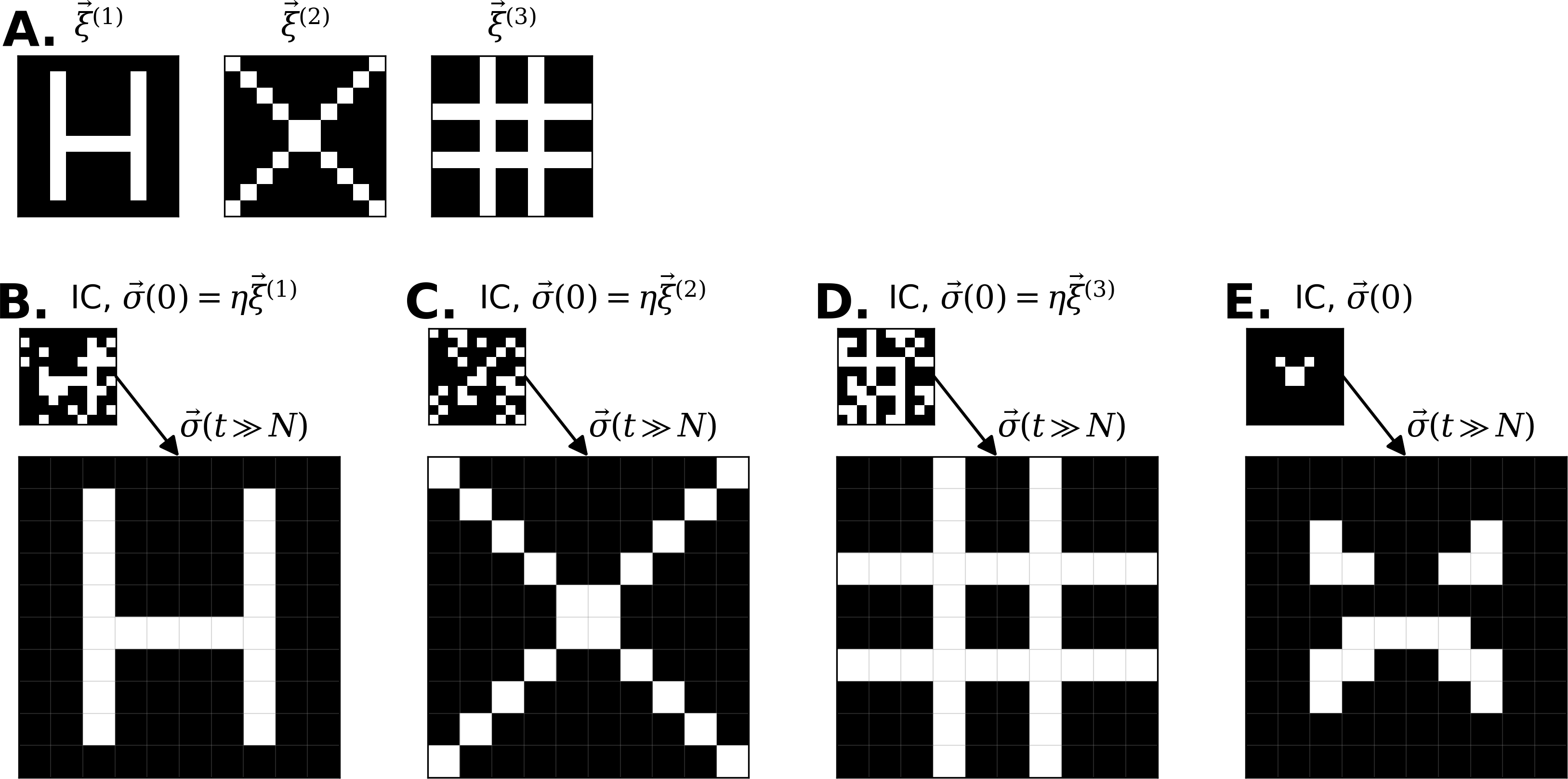}
    \caption{\label{fig:hopfield3mem}
    Network with $N=100$ neurons; states are displayed using the convention  in Fig.~\ref{fig:retrieval}(a).
    (a)~Patterns stored in the connectivity matrix $\mathbf{W}$ using Eq.~\eqref{eq:weighthopfield}.
    (b)--(d)~The initial conditions are the memory patterns with $\alpha=20\%$ of their neurons randomly flipped (top);
    the attractor reached after iterating Eq.~\eqref{eq:hopfielddynamics} either synchronously or asynchronously matches the corresponding memory used as the initial condition (bottom).
    (e) A specific initial condition (top) led to a spurious attractor (hallucination) after asynchronously iterating 10\,MC steps.
    }
\end{figure*}

Note that if $\ctalk\approx0$, Eq.~\eqref{eq:itercrosstalk} reduces to Eq.~\eqref{eq:attractor1mem},
giving $\sigma_i(t+1)=\sigma_i(t)=\sigma_i^*=\spinmemstate[\nu]_i$, as desired.
The term in brackets in Eq.~\eqref{eq:crosstalk} is the overlap, or correlation,
between memories $\mu$ and $\nu$,
$\frac{1}{N}\memstate[\mu]\cdot\memstate[\nu]$.
Correlated memories produce nonzero overlaps, increasing the crosstalk and impairing memory retrieval.
For this reason, the Hopfield model assumes approximately uncorrelated memories,
so that $\ctalk$ remains small and the network can reliably recover stored patterns.
The model can also work with correlated memories, but the network must be trained
using a different approach.\cite{hertzRedesNeurais,peretto}

Because the crosstalk depends only on the stored memories,
Eqs.~\eqref{eq:itercrosstalk} and~\eqref{eq:crosstalk}
can be used to estimate the storage capacity $P_\mmax$,
the maximum number of memories that can be stored
without compromising retrieval.
We define
\begin{equation}
\cctalk=-\spinmemstate[\nu]_i(\ctalk+H_i)\ ,
\end{equation}
and write Eq.~\eqref{eq:itercrosstalk} as
\begin{equation}
\label{crosstalk2}
\sigma_i(t+1)=\Ffunc[\spinmemstate[\nu]_i - \frac{\cctalk}{\spinmemstate[\nu]_i}]
=\Ffunc[\frac{1-\spinmemstate[\nu]_i\cctalk}{\spinmemstate[\nu]_i}]\ ,
\end{equation}
If $\cctalk>1$, the neuron flips even when initialized
in the correct memory state $\spinmemstate[\nu]_i$, making the memory unstable.
Although the network may still converge, the retrieved pattern will contain at least one error.

Assuming random patterns and $N>P\gg1$, the $\cctalk$ term is
a sum of many independent random variables taking values $\pm1$.
By the central limit theorem, we can then approximate it
by a Gaussian variable with zero mean and variance $P/N$.
The error probability $\Perr$ is the probability that $\cctalk>1$,
\begin{align}
    \Perr & =\mathcal{P}(\cctalk>1)=\sqrt{\dfrac{N}{2\pi P}}\int_{1}^{\infty}\!\!e^{-x^2 N/(2P)}{\rm d}x\\
    \Perr&=\frac{1}{2}\left[1-\erf\!\!\left(\sqrt{\frac{N}{2P}}\right)\right]\ ,
    \label{eq:errorprob}
\end{align}
where $\erf\!\!\left(x\right)\equiv(2/\sqrt{\pi})\int_0^x e^{-u^2}du$.

We set $P= P_\mmax$ for  $\Perr=1\%$ in Eq.~\eqref{eq:errorprob} and find
$\erf\!\!\left(\sqrt{N/(2P_\mmax)}\right)=1-2\Perr=0.98$.
The numerical solution of the error function  yields
$\sqrt{N/(2P_\mmax)}\approx1.645$,
leading to the estimate
$P_\mmax\approx0.185N$
for the maximum storage capacity.
Beyond this limit, retrieval errors become frequent.
Even a small number of unstable neurons can trigger avalanches of spin flips,
causing the network to converge to an incorrect or spurious memory, or prevent convergence altogether.
A more conservative criterion requires the probability of error to be 1\% per neuron, that is $\Perr\approx0.01/N$.
In the large $N$ limit, we may use the asymptotic expansion
$1-\erf(x)\sim e^{-x^2}/(x\sqrt{\pi})$, $x\to\infty$, with $x=\sqrt{N/(2P)}$.
Solving for $P$  gives $P_\mmax\sim N/(2\ln\!N)$.

\subsection{Simulation}
\label{sub:simulation}

The iteration of Eq.~\eqref{eq:hopfielddynamics} can be implemented either \textit{synchronously},
with all neurons updated simultaneously at each time step~$t$,
or \textit{asynchronously}, with neurons updated sequentially, one at a time.
In Sec.~\ref{sub:patternstability},
we discussed the conditions under which the asynchronous dynamics converges.
Synchronous dynamics can accelerate convergence toward an attractor.
However, when $H_i=0$, it may lead not only to fixed points,
but also to limit cycles\footnote{A limit cycle of period $q$ is when the network state repeats after $q$ iterations, $\netstate(t+q)=\netstate(t)$.}
with period 2 for symmetric couplings ($W_{ij}=W_{ji}$)
or period 4 for antisymmetric couplings ($W_{ij}=-W_{ji}$).~\cite{peretto}

Asynchronous simulations can be implemented in two main ways.
In the first, a neuron $i$ is selected at each time step $t$,
and the update rule in Eq.~\eqref{eq:hopfielddynamics} is applied.
The selected neuron may be chosen either sequentially or uniformly at random.
In the second approach, each neuron has a constant probability per unit time
of being independently selected for update at any instant~$t$,
potentially allowing more than one neuron update per time step.
Although both methods are straightforward to simulate numerically,
the latter is more naturally suited for hardware implementation.~\cite{hertzRedesNeurais}

The dynamics stops when $\netstate(t)=\netstate(t+1)\equiv\netstate^*$,
indicating that the network has reached an \textit{attractor} $\netstate^*$.
In practice, we may allow for a small tolerance
to account for the probability of error per neuron $\Perr$
or for the presence of cycles.
In this case, the stopping criterion can be relaxed to
$\dHamm[\netstate(t+1),\netstate(t)]\leq \varepsilon N$,
where $\varepsilon\sim\Perr$ is a small parameter and $\dHamm$
is the Hamming distance.\footnote{The Hamming distance $\dHamm[\vec{x},\vec{y}]$
is the number of components in which the vectors $\vec{x}$ and $\vec{y}$ differ.}
Thus, $\netstate(t+1)\approx\netstate^*$.

The existence of attractors  allows the network to correct errors and associate error-prone or mixed
inputs with a well-defined output.
This behavior is a consequence of the nonlinearity of $\Ffunc$. If $\Ffunc$ were linear, then
some linear combination of inputs, say $\netstate(0)=0.3\vec{I}_1+0.7\vec{I}_2$,
would be 
mapped into a linear combination of outputs, $\netstate^*=0.3\vec{O}_1+0.7\vec{O}_2$.
However, the step function forces the network to a \textit{decision}.
Usually, it converges to the nearest attractor, so iterating a Hopfield network
under these conditions would most likely lead to~\cite{Hopfield1982} $\netstate^*\approx\vec{O}_2$.

\subsection{Extensions and biological relevance}
\label{sub:biology}

The resemblance between Eq.~\eqref{eq:hopfielddynamics}
and the equation of state for a magnet (see Eq.~(S-2) in the Supplementary Material)
is no coincidence.
The update rule in Eq.~\eqref{eq:hopfielddynamics}
is often described as a zero-temperature dynamics:
the mean-field solution of the Ising model at zero temperature is~\cite{tragtenbergYokoi}
$m=\sign(Jm+H)$,
which is  the same form as the dynamics adopted here.
Noise can be introduced by replacing the sign function with a hyperbolic tangent,
$\Ffunc(x)\to\tanh\!\left(x/T\right)$,
where $T$ acts as a temperature-like parameter.
The dynamics then becomes stochastic, provided it satisfies detailed balance
in the same spirit as MC methods.~\cite{Hopfield1984,peretto,hertzRedesNeurais,barkemaMC}
More generally, the sign function can be replaced by any bounded,
continuous-valued function --- a sigmoid being the natural choice.
Common choices include~\cite{hertzRedesNeurais,kinouchiDream} $\Ffunc[x]=1/(1+e^{-x})$
and $\Ffunc[x]=x/(1+|x|)$.

At low temperatures, stochastic noise can actually help convergence to stored memories:
small fluctuations may push the network state out of local energy minima
through occasional updates that increase the energy function.
At high temperatures, however, the dynamics becomes excessively random
and the network loses its ability to retrieve memories altogether.
This behavior is  analogous to what is observed in ferromagnets and spin glasses.

We can also write the dynamics in a form that admits
a natural neurophysiological interpretation.~\cite{Hopfield1984}
Neurons emit voltage spikes when stimulated (Fig.~\ref{fig:biology}), and this
can be quantified by the number of spikes per unit time $x_i=(\sigma_i+1)/2$
emitted as a function of the input intensity $h_i$, illustrated in Fig.~\ref{fig:biology}B.
The simplest dynamics that yields this input-output relation is given by
\begin{equation}
\label{eq:hopfielddyn2}
\tau_i \dfrac{{\rm d}\sigma_i}{{\rm d}t}=-\sigma_i+\Ffunc[\sum_{j=1}^{N}W_{ij}\sigma_j(t) + H_i]\ ,
\end{equation}
where $\sigma_i(t)$ can now be a continuous variable rather than a binary state,
and $\tau_i$ is the time constant of the membrane\footnote{A membrane is a thin layer
    surrounding the neuron across which ionic and molecular exchanges generate and propagate
    electrical signals known as spikes or action potentials on a timescale of milliseconds.~\cite{GerstnerBook2014}},
reflecting its passive electrical properties, such as capacitance and resistance.~\cite{GerstnerBook2014}
Note that the dynamics that we proposed earlier, Eq.~\eqref{eq:hopfielddynamics},
is simply Eq.~\eqref{eq:hopfielddyn2} integrated using the Euler method with a time step $\Delta t=\tau_i$,
while measuring time in units of $\tau_i$. This correspondence requires that the membrane time scale, $\tau_i$,
is much faster than the firing rate time scale described by Eq.~\eqref{eq:hopfielddynamics}.
The function $\Ffunc$ models the neuron's input--output relation,
capturing thresholding and saturation effects that can be measured experimentally.

In an equivalent formulation  the dynamics is written in terms of the local fields,~\cite{hertzRedesNeurais}
\begin{equation}
\label{eq:hopfielddynIF}
\tau_i \dfrac{{\rm d}h_i}{{\rm d}t}=-h_i+ H_i+\sum_{j=1}^{N}W_{ij}\Ffunc[h_j]\ ,
\end{equation}
where $h_i$ [Eq.~\eqref{eq:localfield}] is the effective local field on neuron $i$.
Here the inputs are first integrated into $h_i$, then mapped to a firing rate by $\Ffunc$
on the membrane of the postsynaptic neuron.
This separation between input integration and output generation mirrors the biological
distinction between membrane potential dynamics and spike generation,
forming the conceptual basis of  integrate-and-fire neuron models.~\cite{GerstnerBook2014,Shimoura2021Conn}
Such models have been shown to reproduce biological features of over 600 neurons.~\cite{Teeter2018GLIF,Trinh2023hMC}
All these formulations of the Hopfield dynamics 
share the same attractor structure provided $\mathbf{W}$ is invertible.~\cite{Pineda1988}

The fact that attractor dynamics carries over to these alternative formulations
gives the model a greater degree of biological fidelity (Fig.~\ref{fig:biology}).
Although the original Hopfield model assumes fixed synaptic weights $W_{ij}$,
biological networks continuously modify their synapses through activity-dependent plasticity.
One key mechanism is spike-timing-dependent plasticity, where 
synapses are strengthened when presynaptic firing precedes postsynaptic firing, and weakened otherwise,
as observed in hippocampal and cortical slices.~\cite{Markram1996Syn}
Spike-timing-dependent plasticity implements a temporally refined version of the Hebbian learning rule,~\cite{Shimoura2021Conn}
offering a biologically realistic way to update $W_{ij}$ from ongoing neural activity.

Taken together, activity-dependent plasticity and continuous-time dynamics suggest
that the Hopfield model captures the mechanisms  underlying associative memory in animals.
Empirically observed memory and pattern matching can thus be understood as dynamical processes
unfolding in a high-dimensional state space,
where memories correspond to stable attractors and cognitive processes to trajectories between them.
This perspective has influenced a broad range of biological theories,
from working memory and pattern completion to recurrent neural network architectures.~\cite{RollsTreves1999Book,RollsTreves2024}

\section{Example problems}\label{sec:examples}

We present several  problems with solutions
that can be  adapted for classroom use.
For convenience, the problems are organized by topic:
computational physics, dynamical systems, linear algebra,
and statistical physics.

\subsection{Computational Physics}
\label{sub:compphys}

\begin{figure}[t!]
\centering
    \includegraphics[width=1\linewidth]{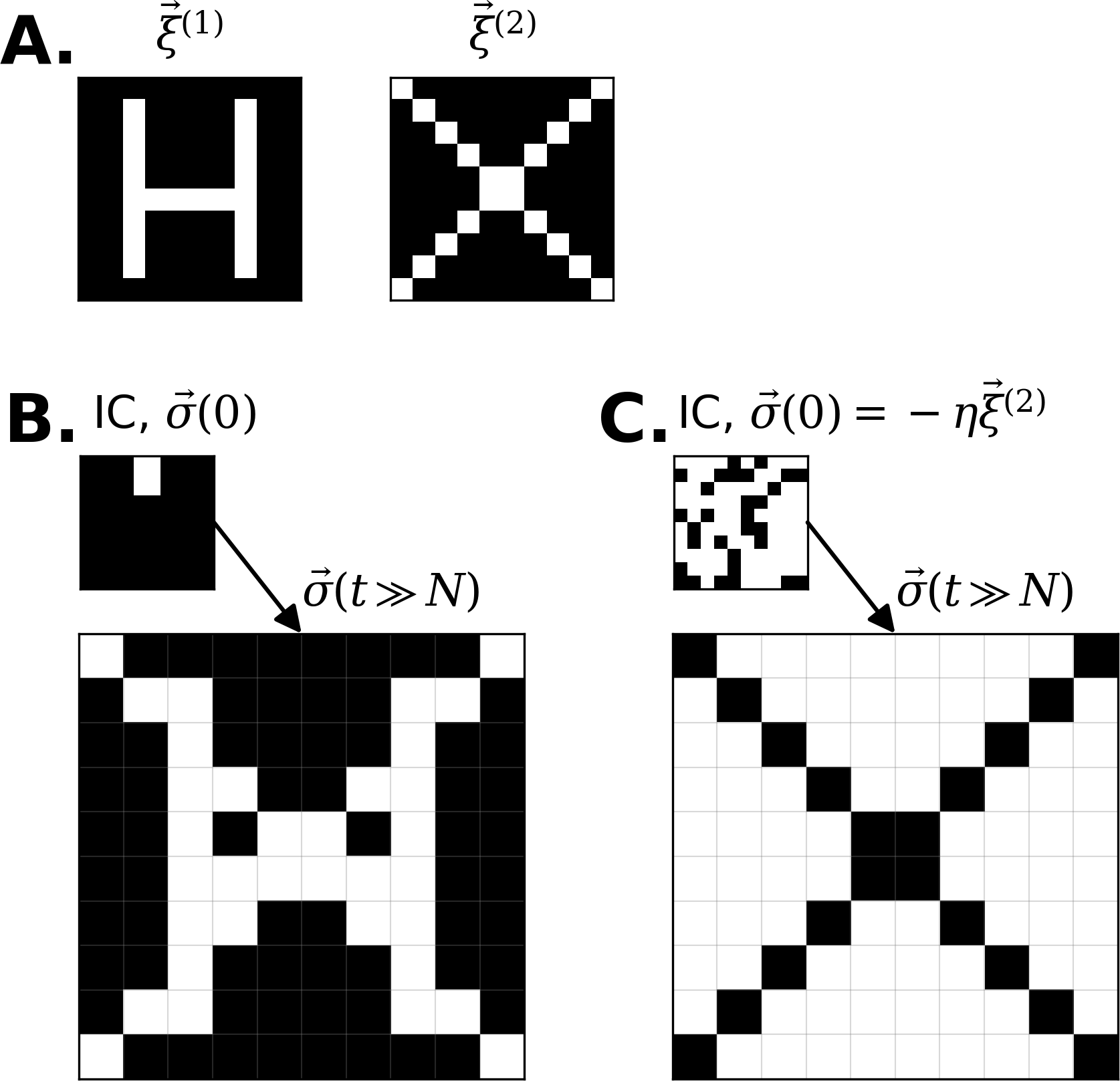}
    \caption{\label{fig:hopfield2hall}
    Network with $N=100$ neurons and two stored memories; states are displayed using the convention in Fig.~\ref{fig:retrieval}(a).
    (a)~Patterns stored in the connectivity matrix $\mathbf{W}$ using Eq.~\eqref{eq:weighthopfield}.
    (b)~A specific initial condition (top) led to a spurious attractor after synchronously iterating for 15\,MC steps.
    (c)~The initial condition was the flipped ``X'' pattern with $\alpha=30\%$ of its neurons randomly flipped (top);
    the attractor reached after iterating Eq.~\eqref{eq:hopfielddynamics},
    either synchronously or asynchronously, matches the flipped memory used as the initial condition (bottom).
    }
\end{figure}

The most direct application of the Hopfield model is in the context of teaching computational physics.
The basic Hopfield model theory is deterministic so it should be easier to grasp than, for example,
the Metropolis  algorithm. 
Many systems in physics can be simulated similarly
to the Hopfield model.
These systems include
contact processes,~\cite{dickmanBook1999,OliveiraDickman2005,noneqPhaseTrans2008,Girardi2025CPSim}
algorithms for MC simulations,~\cite{barkemaMC}
and more complex biologically-motivated neural network simulators.~\cite{NESTSim2007,Shimoura2021Conn}

The implementation of a basic Hopfield network from scratch can be a project for students.
Instructors can also give part of our   code [see 
the Supplementary Material] and suggest particular studies or modifications.
In the following, we propose several problems for students who have a working code.

\exampleproblem\label{prob:attractors} Create a network of $N=100$ neurons and store three memories,
$\{\memstate[1],\memstate[2],\memstate[3]\}$. Make sure the memories are reasonably uncorrelated,
i.e., $\overlapmem[\memstate[\mu],\memstate[\nu]]\approx0$, and easily distinguishable by visual inspection
(e.g.,   use patterns such as ``H'', ``X'' and ``\#'').
Take the first memory $\memstate[1]$, flip a fraction $\alpha$ of its neurons and use it as the initial configuration.
Study the convergence of the network to $\memstate[1]$ as $\alpha$ is increased from 0 to 1 in steps of 0.1.
At what value $\alpha=\alpha_c$ does the convergence become unreliable? Repeat  using $\memstate[2]$ and $\memstate[3]$.
Are the values of $\alpha_c$ for each memory equal?

To flip a fraction $\alpha$ of the memory states, we can construct an identity-like matrix $\eta$ where a random fraction $\alpha$
of its diagonal entries are $-1$. Then, $\netstate(0)=\eta\memstate[1]$ can be used as the initial condition.
An example of the output for $\alpha=0.2$ is shown in Figs.~\ref{fig:hopfield3mem}(a)--(d).

We suggest fixing the seed of the random number generator before running the program
so that results are reproducible. It is  better to use asynchronous updates,
because convergence is guaranteed below the storage capacity.

\exampleproblem\label{prob:antimem} Repeat Prob.~\ref{prob:attractors} starting from the flipped memories,
$-\memstate[1]$, $-\memstate[2]$, and $-\memstate[3]$, with a fraction $\alpha$ of neurons reversed.
Does the network converge to the corresponding attractor $\memstate$ or to its flipped 
versions, $-\memstate$? Your result shows that flipped attractors are also valid memories [e.g., see Fig.~\ref{fig:hopfield2hall}(c)].

\exampleproblem\label{prob:spuriousmem} During your exploration of initial conditions in Prob.~\ref{prob:attractors},
did you find an attractor that was not in the set of memories, $\{\memstate[1],\memstate[2],\memstate[3]\}$? 
If so, for what $\alpha$ values did this happen?
Construct initial conditions by mixing parts of two or more memories so that the network converges
to these spurious attractors. Spurious attractors can be found by either asynchronous [see Fig.~\ref{fig:hopfield3mem}(e)]
or synchronous [see Fig.~\ref{fig:hopfield2hall}(b)] dynamics.

\exampleproblem\label{prob:overlap}Consider the code for memory retrieval
from an initial condition using asynchronous updates (see the Supplementary Material).
Modify it to calculate and return the overlap of the network state $\netstate(t)$
with every memory $\memstate[\mu]$ at each time step, $\overlapmem[\netstate(t),\memstate[\mu]]$.
Plot all the overlaps versus the time for a network with $N=100$ neurons and three memories.
Use this plot and repeat Probs.~\ref{prob:attractors}, \ref{prob:antimem} and~\ref{prob:spuriousmem}
with $\alpha=0.2$.
It is easier to track the convergence by the overlap than by visual inspection.
What is the value of the overlap when the network converges to a memory, an antimemory,
or a spurious attractor?

\begin{figure}[t!]
\centering
    \includegraphics[width=1\linewidth]{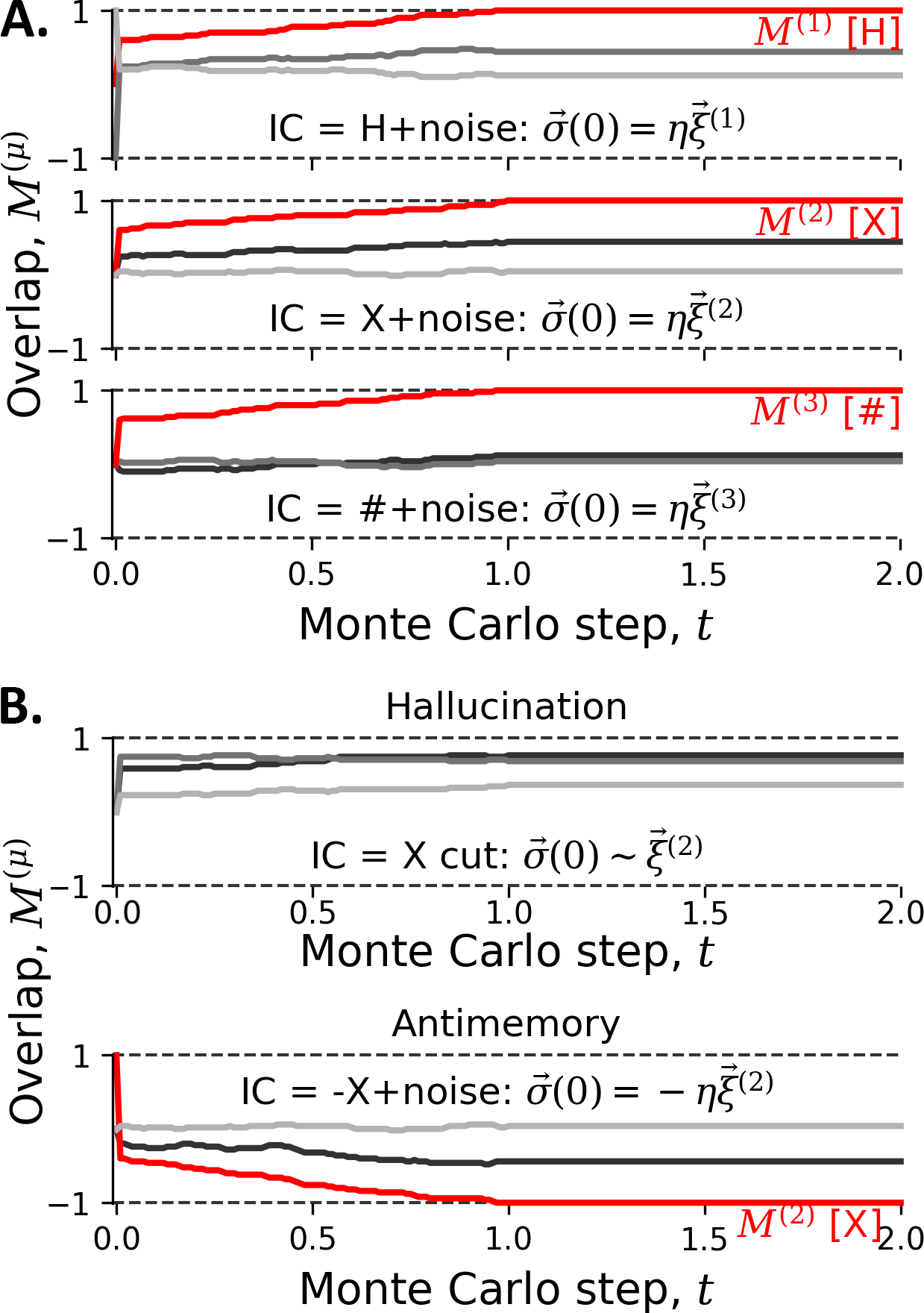}
    \caption{\label{fig:dynamicsoverlap}
    (a)~The overlap goes to $\overlapmem=+1$ when the network converges to one of the memories $\memstate[\mu]$;
    The initial conditions and attractors are shown in Fig.~\ref{fig:hopfield3mem}.
    (b)~The overlap goes to $\overlapmem=-1$ when the network converges to an antimemory $-\memstate[\mu]$ (bottom),
    or remains trapped within $(-1,1)$ when the network converges to an spurious attractor (top); the initial conditions and attractors are shown
    in Fig.~\ref{fig:hopfield2hall}.
    }
\end{figure} 

The code for storing memories in the matrix $\mathbf{W}$ is in the Supplementary Material.
A solution for Prob.~\ref{prob:overlap} is 
in the function \verb|iterate_hopfield_sequential| in the Supplementary Material. 
An example of what can be seen in each situation is shown in Fig.~\ref{fig:dynamicsoverlap}:
memories converge to $\overlapmem=+1$, antimemories converge to $\overlapmem=-1$,
and spurious attractors converge to a value in between.

\exampleproblem\label{prob:energy}Modify the code given in the Supplementary Material
to compute the energy function of the system at each time step.
Note that all thresholds are zero. Create a network with $N=100$ neurons and two memories, $\{\memstate[1],\memstate[2]\}$.
Choose $\memstate[1]$ to be vertical stripes (first stripe in the left is $-1$, second is $+1$, etc), and $\memstate[2]$ to be horizontal stripes
(first stripe from the top is $+1$, second is $-1$, etc).
The overlap between these memories is zero.
Set the entire network to $-1$, except for the first two vertical stripes of memory $\memstate[1]$ as the initial condition.
Run the dynamics and record the energy as a function of time.
Do the same for $\memstate[2]$, but with an initial condition set to $+1$, except for the first two horizontal stripes of $\memstate[2]$.
Compare the evolution of the energy of these two patterns. Can you distinguish the patterns by the energy? Why or why not?

\begin{figure}[t!]
\centering
    \includegraphics[width=1\linewidth]{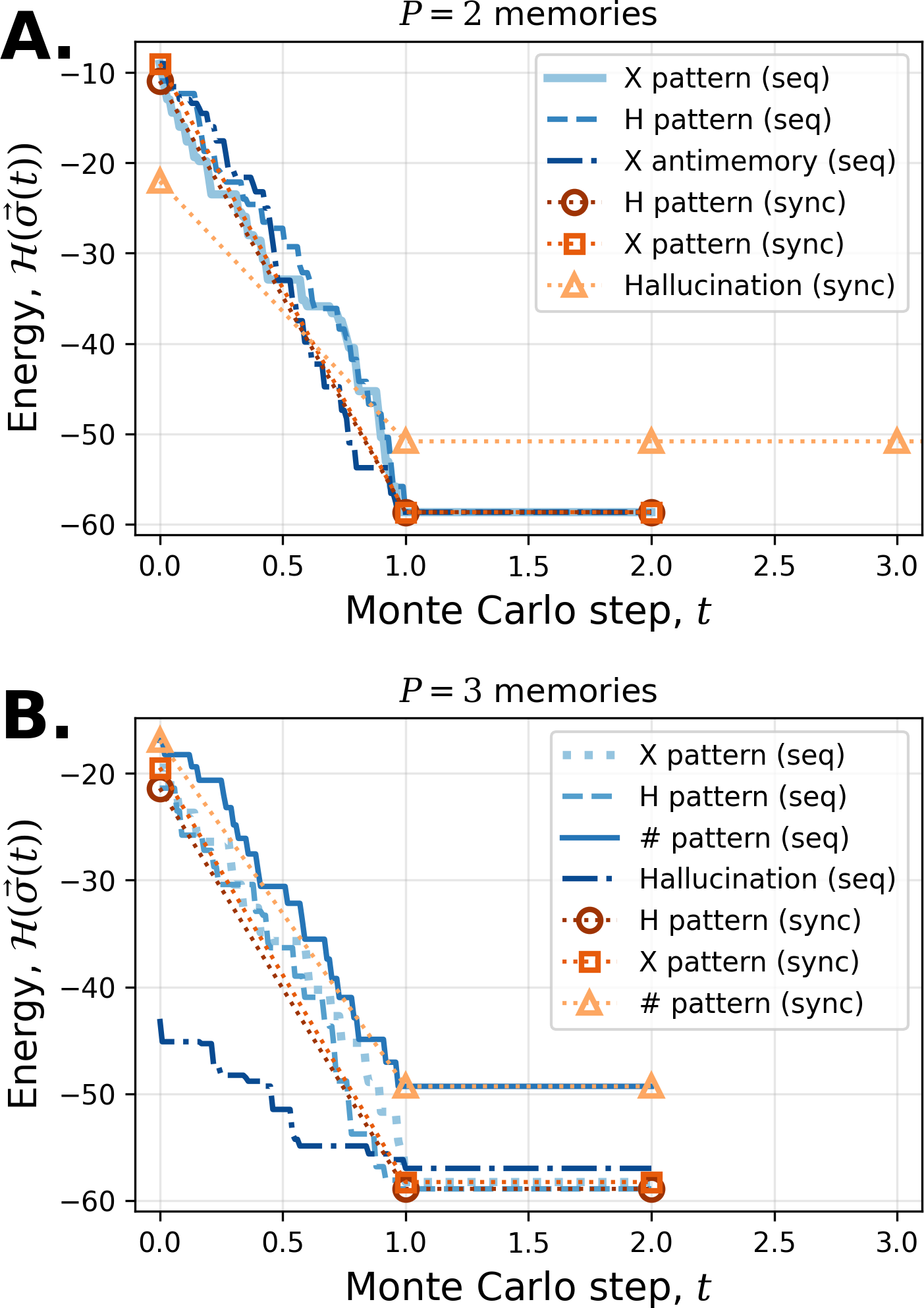}
    \caption{\label{fig:dynamicsenergy}
    (a)~The evolution of the energy during each retrieval of Fig.~\ref{fig:hopfield2hall}'s memories.
    (b)~The evolution of the energy during each retrieval of Fig.~\ref{fig:hopfield3mem}'s memories.
    Hallucinations (attractors that were not purposefully encoded in the Hamiltonian) are evidence of local minima. $N=100$ neurons.
    }
\end{figure}

Both patterns have the same energy, because they both have the same number of activated and deactivated neurons.
Thus, it is not possible to distinguish between both patterns  by looking only at the energy, even if 
both patterns are completely uncorrelated. For example, in Fig.~\ref{fig:dynamicsenergy}
we show the evolution of the energy of the network while retrieving the patterns from Figs.~\ref{fig:hopfield3mem},
and~\ref{fig:hopfield2hall}. The ``X'' and ``H'' patterns have the same energy.

\exampleproblem The storage capacity of a network is $\approx 0.138 N$.
Create a network with $N=100$ neurons. Add ten random memories and run the dynamics from a few initial conditions.
Use the energy and overlap plots to determine whether the dynamics converged to a memory (convergence requires 
that the overlap is $M=\pm1$ between a memory and the final network state).
Next add ten more random memories, run the dynamics again from another set of initial conditions,
and check convergence from the energy and overlap plots. How does this plot compare with the first one?
Keep adding more memories and checking the overlap and memory plots. Is the network stable?
The network becomes saturated as more memories are added. Eventually, unstable patterns can emerge, and avalanches
of flipping neurons can occur.

\subsection{Dynamical Systems}
\label{sub:dynsys}

The convergence and stability problems presented in
Sec.~\ref{sec:hopfield} can be discussed in nonlinear dynamics courses.

\exampleproblem\label{prob:attractorstable} Consider a network of $N$ neurons with $P=1$.
Represent the network state as a column vector and show that the memory $\memstate$
is an attractor if $H_i\in(-1,1)$.

This problem was solved in deriving Eq.~\eqref{eq:attractor1mem}. We can define the local field
column vector as $\vec{h}(t)=\mathbf{W}\netstate(t)+\vec{H}$, such that $\netstate(t+1)=\Ffunc[\vec{h}(t)]$
is the sign function. $\vec{H}$ is a column vector with the threshold for each neuron, and $\mathbf{W}$
is given by Eq.~\eqref{eq:weightmatrix}. We take $\netstate(t)=\memstate$ and need to retrieve it by a single iteration,
\begin{align}
\netstate(t+1)&=\Ffunc[\vec{h}(t)]=\Ffunc[\mathbf{W}\memstate+\vec{H}] \\
&=\Ffunc[\frac{1}{N}\memstate\,\transpose{\memstate}\memstate+\vec{H}]
=\Ffunc[\memstate+\vec{H}]\ ,\label{eq:1memstabexer}
\end{align}
where we used the overlap definition as the scalar product,
\begin{equation}
N\overlapmem[\memstate,\memstate\,]=\transpose{\memstate}\memstate=N\ .
\end{equation}
We recall that $\spinmemstate_i=\pm1$, and  write Eq.~\eqref{eq:1memstabexer} as
\begin{equation}
\Ffunc[\spinmemstate_i+H_i]=\left\{\begin{array}{ll}
+1 & \text{, if } \spinmemstate_i\geq -H_i\\
-1 & \text{, if } \spinmemstate_i< -H_i
\end{array}\right\}=\spinmemstate_i\ ,
\end{equation}
because $\spinmemstate_i=+1>-H_i$
and  $\spinmemstate_i=-1<-H_i$ for $H_i\in(-1,1)$.

\exampleproblem\label{prob:crosstalk} Consider a network of $N$ neurons with $P$ memories
and all thresholds $H_i=0$.
Represent the network state as a column vector and show that the memory $\memstate[\nu]$
is an attractor if $\overlapmem[\memstate[\mu],\memstate[\nu]]\approx0$,
i.e., the overlap (or correlation) between memory $\nu$ and any other memory is negligible.

As in Prob.~\ref{prob:attractorstable},
we need to evaluate the local field $\vec{h}(t)$ applied to the state $\netstate(t)=\memstate[\nu]$,
\begin{align}
\netstate(t+1)&=\Ffunc[\vec{h}(t)]=\Ffunc[\mathbf{W}\memstate[\nu]]\\
&=\Ffunc[\frac{1}{N}\sum\limits_{\mu=1}^{P}\memstate[\mu]\transpose[]{\memstate[\mu]}\memstate[\nu]] \\
&=\Ffunc\Bigg(\frac{1}{N}\memstate[\nu]\transpose[]{\memstate[\nu]}\memstate[\nu]\nonumber\\
&\qquad +\sum\limits_{\substack{\mu=1\\\mu\neq\nu}}^{P}\memstate[\mu]\left(\frac{1}{N}\transpose[]{\memstate[\mu]}\memstate[\nu]\right)\Bigg)\\
&=\Ffunc[\memstate[\nu]
+\sum\limits_{\substack{\mu=1\\\mu\neq\nu}}^{P}\memstate[\mu]\overlapmem[\memstate[\mu],\memstate[\nu]]]\ .\label{eq:2memstabexer}
\end{align}
If $\left|\overlapmem[\memstate[\mu],\memstate[\nu]]\right|< \varepsilon/P\ll1$ for all $\mu\neq\nu$, 
and $0<\varepsilon<1$, then Eq.~\eqref{eq:2memstabexer}
reduces to $\netstate(t+1)=\Ffunc[\memstate[\nu]+\orderof{\varepsilon}]=\memstate[\nu]$.
Because $\memstate[\nu]$ is retrieved in one iteration, it is an attractor of the dynamics.

\exampleproblem\label{prob:basins} Create the weight matrix for an $N=10$ network
and store (a) 1, (b) 2, and (c) 3 memories in it using the Hebb rule,
Eq.~\eqref{eq:weighthopfield}.
Compute the Hamiltonian, Eq.~\eqref{eq:energygeneral},
with $H_i=0$ for all $2^N=1024$ possible spin configurations,
assigning a unique label to each state.
Use each configuration as an initial condition and iterate the Hopfield dynamics
until convergence (retrieval function \verb|get_memory_async|).
Identify which memory (or anti-memory) each initial condition converges to.
Then, use the  function \verb|sort_basins| to reorganize the states
according to their attractor basins.
Finally, plot the sorted Hamiltonian versus the reordered state labels,
coloring each state according to its basin of attraction.
Compare the basin organization for the cases of one, two, and three stored memories.

The basin of attraction of a memory is the set of initial conditions that converge to it under the network dynamics.
Its structure is directly related to the convergence properties discussed in Sec.~\ref{sub:compphys}
and to the geometry of the Hamiltonian landscape.
Figure~\ref{fig:energylandscape} illustrates a convenient ordering of states
that exposes the energy wells associated with different attractors.
In this problem, we use brute force on a small network so that all $2^N$ states can be analyzed explicitly,
providing a complete visualization of the attractor basins and their corresponding energy landscape.

\subsection{Linear Algebra}
\label{sub:linalgebra}

Linear algebra is often one of the first mathematical tools
encountered by undergraduate physics students, yet its role in research
is not usually immediately apparent. Concepts such as vectors, dot products, and matrices are typically taught in abstract 
settings or applied to canonical examples like rotations and eigenvalue problems.
Encoding and recalling information naturally emerge from vector operations in the Hopfield model.
Thus, the problems in Sec.~\ref{sub:dynsys} can also be used in linear algebra classes. Another example is given in Prob.~\ref{prob:ex}.

\exampleproblem \label{prob:ex} Show that the weights, Eq.~\eqref{eq:weighthopfield}, form a symmetric matrix
given by the sum of external products between the memory states $\memstate[\mu]$, each of which is written as a column vector.

Problem~\ref{prob:ex} can be introduced by    discussing the importance of vector and matrix algebra
in artificial intelligence. The instructor can derive the $W_{ij}$ from Eqs.~\eqref{eq:overlap} and~\eqref{eq:energydef},
and ask students to make this simple demonstration,
    \begin{align}
        \mathbf{W}&=\dfrac{1}{N}
        \left[
            \begin{array}{cccc}
                \sum\limits_{\mu=1}^P\spinmemstate[\mu]_1\spinmemstate[\mu]_1& \sum\limits_{\mu=1}^P\spinmemstate[\mu]_1\spinmemstate[\mu]_2 & \dots  & \sum\limits_{\mu=1}^P\spinmemstate[\mu]_1\spinmemstate[\mu]_N\\
                \sum\limits_{\mu=1}^P\spinmemstate[\mu]_2\spinmemstate[\mu]_1& \sum\limits_{\mu=1}^P\spinmemstate[\mu]_2\spinmemstate[\mu]_2 & \dots  & \sum\limits_{\mu=1}^P\spinmemstate[\mu]_2\spinmemstate[\mu]_N\\
                \vdots                                                       &              \vdots                                           & \ddots & \vdots              \\
                \sum\limits_{\mu=1}^P\spinmemstate[\mu]_N\spinmemstate[\mu]_1& \sum\limits_{\mu=1}^P\spinmemstate[\mu]_N\spinmemstate[\mu]_2 & \dots  & \sum\limits_{\mu=1}^P\spinmemstate[\mu]_N\spinmemstate[\mu]_N\\
            \end{array}
            \right] \\
        &=\dfrac{1}{N}\sum_{\mu=1}^P
        \left[
            \begin{array}{cccc}
                \spinmemstate[\mu]_1\spinmemstate[\mu]_1& \spinmemstate[\mu]_1\spinmemstate[\mu]_2 & \dots  & \spinmemstate[\mu]_1\spinmemstate[\mu]_N\\
                \spinmemstate[\mu]_2\spinmemstate[\mu]_1& \spinmemstate[\mu]_2\spinmemstate[\mu]_2 & \dots  & \spinmemstate[\mu]_2\spinmemstate[\mu]_N\\
                \vdots                                  &     \vdots                               & \ddots &   \vdots\\
                \spinmemstate[\mu]_N\spinmemstate[\mu]_1& \spinmemstate[\mu]_N\spinmemstate[\mu]_2 & \dots  & \spinmemstate[\mu]_N\spinmemstate[\mu]_N\\
            \end{array}
            \right]\\
        &=\dfrac{1}{N}\sum_{\mu=1}^P\memstate[\mu]\transpose[]{\memstate[\mu]}\ .\label{eq:weightmatrix}
        \end{align}
Symmetry follows from the commutativity of the multiplication between $\spinmemstate[\mu]_i$ and $\spinmemstate[\mu]_j$.

A follow-up question can deal with the retrieval and stability of memories, treating the memory states as vectors and
the couplings as a matrix.

\subsection{Statistical Physics}

\begin{figure}[t!]
\centering
    \includegraphics[width=1\linewidth]{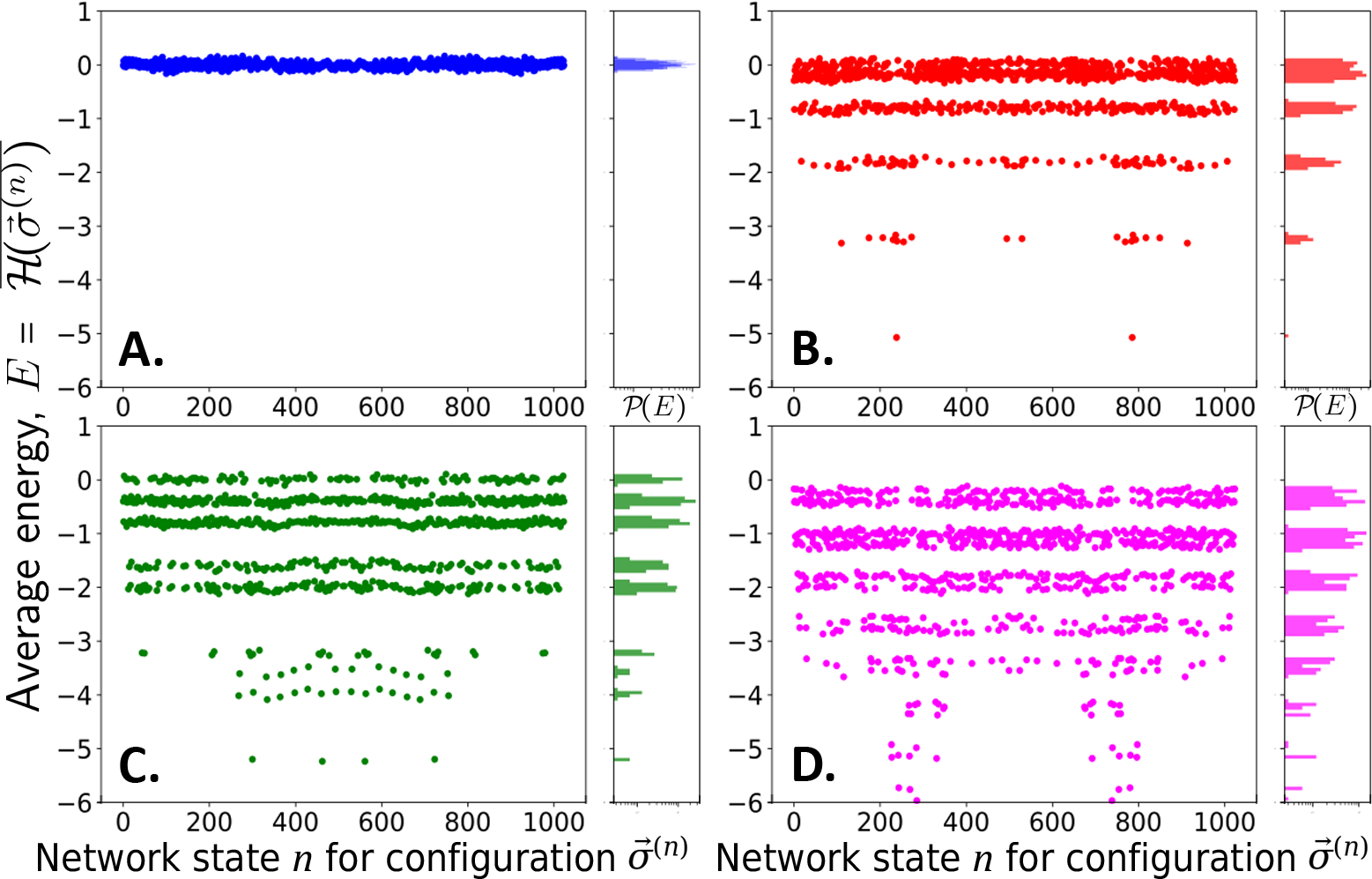}
    \caption{\label{fig:avgenergymemory}
    The  energy averaged over 1000 realizations of the interaction matrix $\mathbf{W}_{\rm eff}$
    for every $\netstate[n]$ state in a network with $N=10$.
    $\mathbf{W}_{\rm eff}=\mathbf{W}_{\rm rand}+\mathbf{W}$, where $\mathbf{W}_{\rm rand}$ is a random matrix
    sampled from a  Gaussian distribution, and $\mathbf{W}$ is the Hebbian 
    matrix, Eq.~\eqref{eq:weighthopfield}.
    The storage capacity is just about one memory (for stable attractors with less than 1\% error).
    (a)~$\mathbf{W}_{\rm eff}$ is  given only by the random part (no memory stored); this average energy
    is equivalent to the spin glass  average energy in Fig.~\ref{fig:isingenergy}.
    (b)~Only a single memory $\memstate$ is encoded in $\mathbf{W}$. It is the same memory for every realization of 
    $\mathbf{W}_{\rm eff}$. The two minima in $\Hfunc$ correspond to $\memstate$ and $-\memstate$.
    The energy levels that arise reflect the discrete nature of the Hamiltonian.
    (c)~The same two memories are encoded in $\mathbf{W}$ for every $\mathbf{W}_{\rm eff}$.
    More energy levels appear because the network is  above the stable storage capacity.
    (d)~The same three memories are encoded in $\mathbf{W}$ for every $\mathbf{W}_{\rm eff}$.
    Many energy levels are now present, making local minima that can trap the dynamics into hallucinations.
    The averaging smooths the random fluctuations and make the minima due to 
    the Hebbian part apparent.}
\end{figure}

Although the Hopfield model is constructed  from the basic ideas of statistical mechanics,
making this connection explicit  is often not straightforward.
Its Hamiltonian has the same structure as those encountered in spin systems,
and temperature plays a role analogous to thermal noise.
Just as high temperature destroys magnetic order in the Ising model,
it also destabilizes memory retrieval in Hopfield networks, 
driving the system toward random configurations.
In both cases, the system undergoes a transition to a disordered, paramagnetic phase
as the temperature increases.

This parallel allows students to transfer familiar concepts from statistical mechanics
to neural computation.
Ideas such as free-energy landscapes, equilibrium states, and phase transitions
can be used almost unchanged to understand how neural networks store and retrieve information
(Fig.~\ref{fig:avgenergymemory}).

\section{Concluding remarks}

The recent rise of AI is  reshaping both scientific practice and the classroom.
Rather than replacing traditional training, it highlights the need for deeper engagement
with the tools that  mediate much of our work. Future physicists will be expected
not only to use computational and AI-based methods, but to understand their foundations,
limitations, and domains of applicability, so that they can adapt and extend them when necessary.

The Hopfield model provides a direct link between simple physics concepts and  AI.
Some examples of real-world problems that can be solved with these networks include
scheduling of truck deliveries or flights, the automatic movement of drills or robot arms, and
the design of chips with minimal wiring. See Ref.~\citenum{hertzRedesNeurais}, Chap.~4, for more details.
These problems can be mapped onto combinatorial problems, such as the traveling salesman (minimizing travel distance),
or graph bipartite (splitting a graph in two parts connected by the minimum amount of edges).
To solve them, we can define cost functions that can be directly mapped on the Hopfield Hamiltonian,
from which the connection weights and thresholds are easily inferred.\cite{hertzRedesNeurais,Rojas1996}
The energy-minimization dynamics then naturally leads to near-optimal solutions.

Physicists bring valuable insights to AI through concepts like energy landscapes and collective behavior,
which are increasingly relevant in a data-driven economy; this growing intersection calls
for curricula that integrate physics with data analysis and learning algorithms
to better prepare students for emerging technological challenges.

\section*{Data availability statement}

The simulations are available in the Supplementary Material below, and also
on the code repository \url{https://github.com/neuro-physics/hopfield-neural-network}.

\section*{Author Declarations and Contributions}

The authors declare no competing interests.

D.D.C.\ and M.H.\ ran simulations and performed calculations.
C.F.V.\ performed calculations.
M.G.-S.\ performed calculations, proposed, and supervised the work.
All authors wrote the manuscript.

\begin{acknowledgments}
M.G.-S.\ acknowledges 
financial support from Fundação de Amparo a Pesquisa e Inovação do Estado de Santa Catarina (FAPESC),
Edital 21/2024 (Grant n.~2024TR002507) and from
Conselho Nacional de Desenvolvimento Científico e Tecnológico (CNPq Grant n.~302102/2025-6).
This research is supported by INCT-NeuroComp
(CNPq Grant 408389/2024-9).
D.D.C.~and M.H.~thank partial financial support by CNPq-Brazil,
and C.F.V.~thanks partial financial support from CAPES-Brazil.
\end{acknowledgments}

\bibliographystyle{myaipnumbibstyle} 

%

\clearpage

\appendix
\section*{Supplementary material}
\makeatletter
\setcounter{figure}{0}
\setcounter{equation}{0}
\setcounter{section}{0}
\renewcommand{\thesection}{\arabic{section}}
\renewcommand{\thesubsection}{\thesection.\arabic{subsection}}
\renewcommand{\theequation}{S-\arabic{equation}}
\renewcommand{\thefigure}{S\arabic{figure}}
\renewcommand{\@seccntformat}[1]{\csname the#1\endcsname\quad}
\makeatother

\section{Statistical Physics: Ising ferromagnet and spin glass}

\subsection{Mean-field theory of ferromagnets}

We start from a system that is usually approached in standard introductory
courses in undergraduate statistical physics: the Ising paramagnet made of $N$ spins.
Its Hamiltonian is a function of the system microscopic configuration described by the column
vector $\netstate=\transpose{[\sigma_1\ \cdots\ \sigma_N]}$,
(${\rm T}$ for transpose)
\begin{equation}
    \label{seq:paraHamilton_original}
    \Hfunc[\netstate]=-H\sum_{i=1}^N\sigma_i\ ,
\end{equation}
where $H$ is an external magnetic field in suitable units (\textit{e.g.},
in terms of the Bohr magneton and the gyromagnetic constant)
and the spin variables can be $\sigma_i=\pm1$.~\cite{Salinas2001Ingles}
We also loosely refer to $\netstate$ as a ``network state'', although thermodynamically
the precise terminology is ``microscopic state'' or ``microscopic configuration''.
Defining the inverse temperature, $\beta=1/(k_BT)$, where $k_B$ is the Boltzmann constant,
the canonical partition function $Z=[2\cosh(\beta H)]^N$ yields the (dimensionless) equation of state
for the magnetization per spin, $m=\tanh(\beta H)$. This system is not very exciting: the
magnetization is always a direct response to the applied magnetic field:
if $H>0$($<0$), then $m>0$($<0$); otherwise, $H=0$ gives no spontaneous magnetization.

The ferromagnetic state can be created if we replaced $H$ by an effective field,
$H_{\rm eff}=H + J\,m$, where $J>0$ is the result of the spins' own magnetic influence
on one another, giving the equation of state
\begin{equation}
\label{seq:magnetCW}
    m=\tanh(\beta J\,m +\beta H)\ .
\end{equation}
This approach is known as the {\it mean field theory} because we replace the actual interaction of a spin with other spins, by the interaction of a spin with the mean field of the other spins that interact with the given spin. [xx added sentence xx] Now, even if $H=0$, the equation $m=\tanh(\beta J m)$ can have either
one solution, $m=0$, when $\beta J<1$, or three solutions, $m=0$ and $m=\pm m_0$,
when $\beta J>1$. This is because the hyperbolic tangent is a sigmoid function 
bounded within $[-1,+1]$, with a slope of $\tanh(\beta J m)\sim\beta J m+\orderof{m^2}$
near $m=0$. Thus, $\beta_c J=J/(k_B T_c)=1$ is a phase transition with critical
temperature $T_c=J/k_B$. For temperatures $T>T_c$, the system is a paramagnet,
but for $T<T_c$, the system generates spontaneous magnetization and becomes a ferromagnet.

This is the Curie-Weiss mean-field theory of the Ising model,~\cite{Salinas2001Ingles}
and it can be made more rigorous if we started from the Hamiltonian in Eq.~\eqref{seq:paraHamilton_original},
replacing $H$ by the effective field
$H_{\rm eff}=\frac{J}{2N}\sum_j\sigma_j+H$,
\begin{equation}
\label{seq:CWHamilton}
    \Hfunc[\netstate]=-\dfrac{J}{2N}\sum_{i=1}^N\sum_{j=1}^N\sigma_i\sigma_j-H\sum_{i=1}^N\sigma_i\ ,
\end{equation}
where the factor $1/2$ in front of the double sum is there to compensate for the fact that
all terms are being summed twice, because $\sigma_i\sigma_j=\sigma_j\sigma_i$. The Hamiltonian in Eq.~\eqref{seq:CWHamilton}
favors states in which $\sigma_i$ and $\sigma_j$ are aligned (\textit{i.e.}, have the same sign).
The canonical partition function
can be written as~\cite{Carneiro1989,Salinas2001Ingles}
\begin{equation}
    Z=\left(\dfrac{N\beta J}{2\pi}\right)^{\!\!1/2}\int_{-\infty}^{+\infty}\!\!\!\!\!\!\exp\!\left(-\beta N g \right)\,{\rm d}m\ ,
\end{equation}
where $g\equiv g(T,H;m)$ is the Gibbs free energy functional
\begin{equation}
    \label{seq:isingenergy}
    g(T,H;m)=\dfrac{J}{2}m^2-\dfrac{1}{\beta}\ln\!\left(2 \cosh(\beta J m+\beta H)\right)\ .
\end{equation}

\begin{figure}[t!]
    \centering
    \includegraphics[width=1\linewidth]{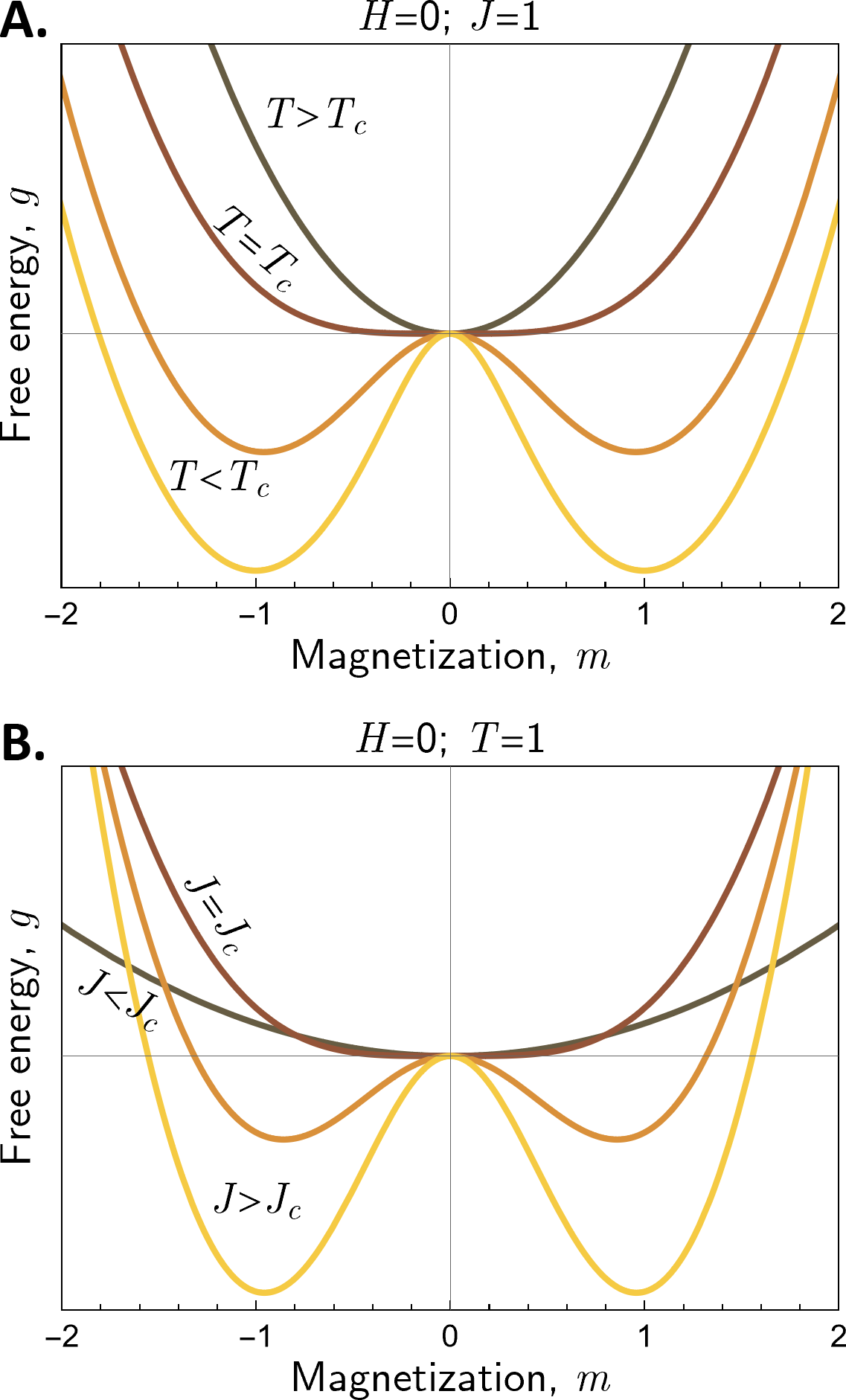}
    \caption{\label{sfig:isingenergy}Ising ferromagnet free energy functional.
    Plots of Eq.~\eqref{seq:isingenergy} using units in which $k_B=1$. The minima of $g$
    are the thermodynamic (observable) states of the system.
    (A)~Break of symmetry as $T$ decreases with fixed $J=1$: the equilibrium $m=0$ ($T>T_c=1$)
    splits into $m=\pm m_0$ for $T<T_c$; both are solutions to the equation of state, Eq.~\eqref{seq:magnetCW}.
    (B)~A similar break of symmetry happens for fixed $T=1$, but changing $J$ around $J_c=1/T=1$.
    Although $J$ is an effective interaction between spins and cannot be changed in real magnets,
    it also controls the thermodynamic (equilibrium) states; this is the fundamental feature
    explored by Hopfield.}
\end{figure}

The minima of $g(T,H;m)$ with respect to $m$ are the thermodynamic states where we can observe the system
[Fig.~\ref{sfig:isingenergy}].
Minimizing Eq.~\eqref{seq:isingenergy}, \textit{i.e.},
finding $m$ such that $\partial g/\partial m=0$,
leads to the magnetization in Eq.~\eqref{seq:magnetCW}.
The usual way to look at $g$ is by keeping $J$ fixed (because it results from the influence
of other spins), and verify that the minima of $g$ respond to changes in $T$.
This is the typical scenario of the critical phase transition,
often termed as ``symmetry breaking'' because the single minimum in $g$ for $T>T_c$
splits into two for $T<T_c$ [Fig.~\ref{sfig:isingenergy}(A)].

If we let go of the fixed $J$ constraint, and think of it as a free parameter,
we see that the minima of $g$ also respond to changes in $J$ when $T$ is fixed [Fig.~\ref{sfig:isingenergy}(B)].
This is the first fundamental principle that explains the functioning of Hopfield networks.
Because $J$ controls the thermodynamic state of the system,
we naively expect that these minima must also be dynamically achievable when the spins
change with time, $\sigma_i(t)$.
Although not in general, this is expected at least in the case where the
(stochastic) dynamics obey detailed balance, or for some
particular deterministic temporal evolution.~\cite{hertzRedesNeurais,peretto}

\subsection{Mean-field theory of spin glasses}

\begin{figure}[t!]
    \centering
    \includegraphics[width=1\linewidth]{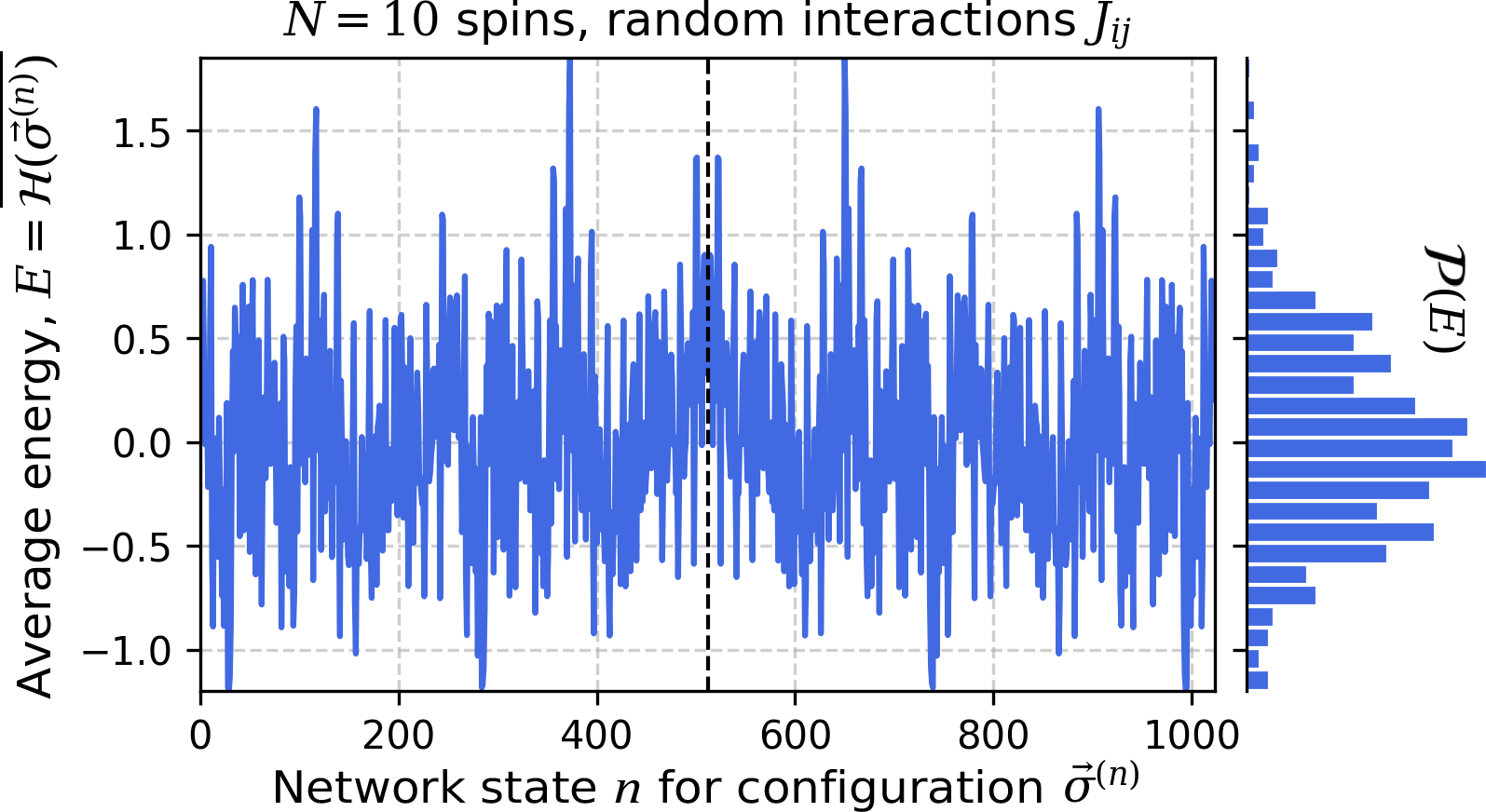}
    \caption{\label{sfig:randomenergy}The energy of a spin glass is shown
    for all $2^N=1024$ microscopic states of a system with $N=10$ spins.
    Every local minimum is a metastable thermodynamic state.
    The curve was obtained by averaging Eq.~\eqref{seq:SKHamilton} over 100 realizations
    of the interaction matrix, $\mathbf{J}$. 
    $J_{ij}$ was sampled from a Gaussian distribution
    with zero mean and unit standard deviation.
    For display purposes, each network configuration $\netstate[n]$ is assigned an index $n$ plotted on the horizontal axis.
    The energy landscape approximately follows a Gaussian distribution (right) with $\avg{E}\approx0$, which motivates
    random-energy models for spin glasses.~\cite{Derrida1980REM,Ruelle1987} 
    The dashed line marks the state $n=2^9=512$. States labeled by $n>512$ correspond
    to spin-flipped configurations, producing the mirror symmetry observed in the plot.
    }
\end{figure}

There is no \textit{a priori}  requirement for  $J$ to be the same  for every pair of spins.
We can imagine that materials  with impurities might result in different interactions
between different pairs of spins. This case is called \textit{quenched} disorder (because it is fixed in time),
and leads to the spin glass Sherrington-Kirkpatrick Hamiltonian~\cite{SherringtonKirk1975}
\begin{equation}
\label{seq:SKHamilton}
    \Hfunc[\netstate]=-\dfrac{1}{2}\sum_{i=1}^N\sum_{j=1}^N\dfrac{J_{ij}}{N}\sigma_i\sigma_j-H\sum_{i=1}^N\sigma_i\ .
\end{equation}
We assume that $J_{ij}=J_{ji}$. If $J_{ij}$ obeys
a distribution with a well-defined average and standard deviation that scales as $1/\sqrt{N}$, the replica method can be applied to Eq.~\eqref{seq:SKHamilton}.\cite{Mezard1987book,SherringtonKirk1975}
Simply put, the replica method consists of evaluating $\Hfunc[\netstate]$
for a given realization (or replica) $\mu$ of  $J_{ij}$, resulting in the Helmholtz free energy
$f^{(\mu)}$;
this procedure is repeated many times to obtain a mean free energy $f=\overline{f^{(\mu)}}$,
where the bar denotes an average over the entire $J_{ij}$ ensemble.  
As an example of a thermodynamic properties that can be derived in a self-consistent way from $f$, we show in Fig.~\ref{sfig:isingenergy}(B) the landscape for
the average energy $E=\overline{\Hfunc[\netstate]}$ over independent realizations
of the $\mathbf{J}$ matrix.

This procedure leads to analytical expressions for the total magnetization per spin, $m$, and for the
Edwards-Anderson~\cite{EdwardsAnderson1975} order parameter, $\qEA$,
\begin{equation}
\label{seq:EAorderpar}
\qEA=\overline{\dfrac{1}{N}\sum_{i=1}^{N}m_i^2}\ ,
\end{equation}
where $m_i=\avg{\sigma_i}$ is the thermal average of spin $i$ (local magnetizatin at site $i$),
and the bar denotes averaging of the entire expression
over the ensemble of $J_{ij}$.~\cite{Mezard1987book} A spin glass is defined as a low-temperature system for which $m=0$,
but $\qEA>0$.~\cite{SherringtonKirk1975}
Although the net magnetization is zero, the system has a random distribution
of locally magnetized domains.

Decreasing the temperature of a spin glass starting from
above the Curie temperature,  will freeze the system in one particular
state $\mu$ out of the many configurations $\{m_i\}$
for the local magnetizations.~\cite{MezardParisi1984}
The overlap of these states is given by
\begin{equation}
\label{seq:spinglassoverlap}
q^{(\mu,\nu)}=\dfrac{1}{N}\sum_{i=1}^{N}m_i^{(\mu)}m_i^{(\nu)},
\end{equation}
where $m_i^{(\mu)}$ is the local magnetization of site $i$ corresponding to replica $\mu$ (obtained from a particular sample of   $J_{ij}$).
By treating the site-related variables as vectors,
$\vec{m}_i^{(\mu)}=\transpose{\left[m_1^{(\mu)}\ \cdots\ m_N^{(\mu)}\right]}$,
the overlap is  the scalar product between two macroscopic states. 

If $\nu=\mu$, $q^{(\mu,\nu)}=q^{(\mu,\mu)}=\qEA$,~\cite{MezardParisi1984}
because the definition in Eq.~\eqref{seq:EAorderpar} should not change from one realization
of the $J_{ij}$ to the other,
so the averaging over the quenched disorder can be neglected.~\cite{Mezard1987book}
We will see that an analogous overlap can  be defined for the Hopfield model, and is crucial
for deriving the Hopfield energy function, and verifying the convergence
of the system to an attractor.~\cite{peretto} Spin glasses  are of much interest in current research in statistical physics, \cite{SherringtonKirk1975,Mezard1987book} but will not be discussed here.

\section{Deriving the Hopfield energy function}

\begin{figure}[b!]
\centering
    \includegraphics[width=1\linewidth]{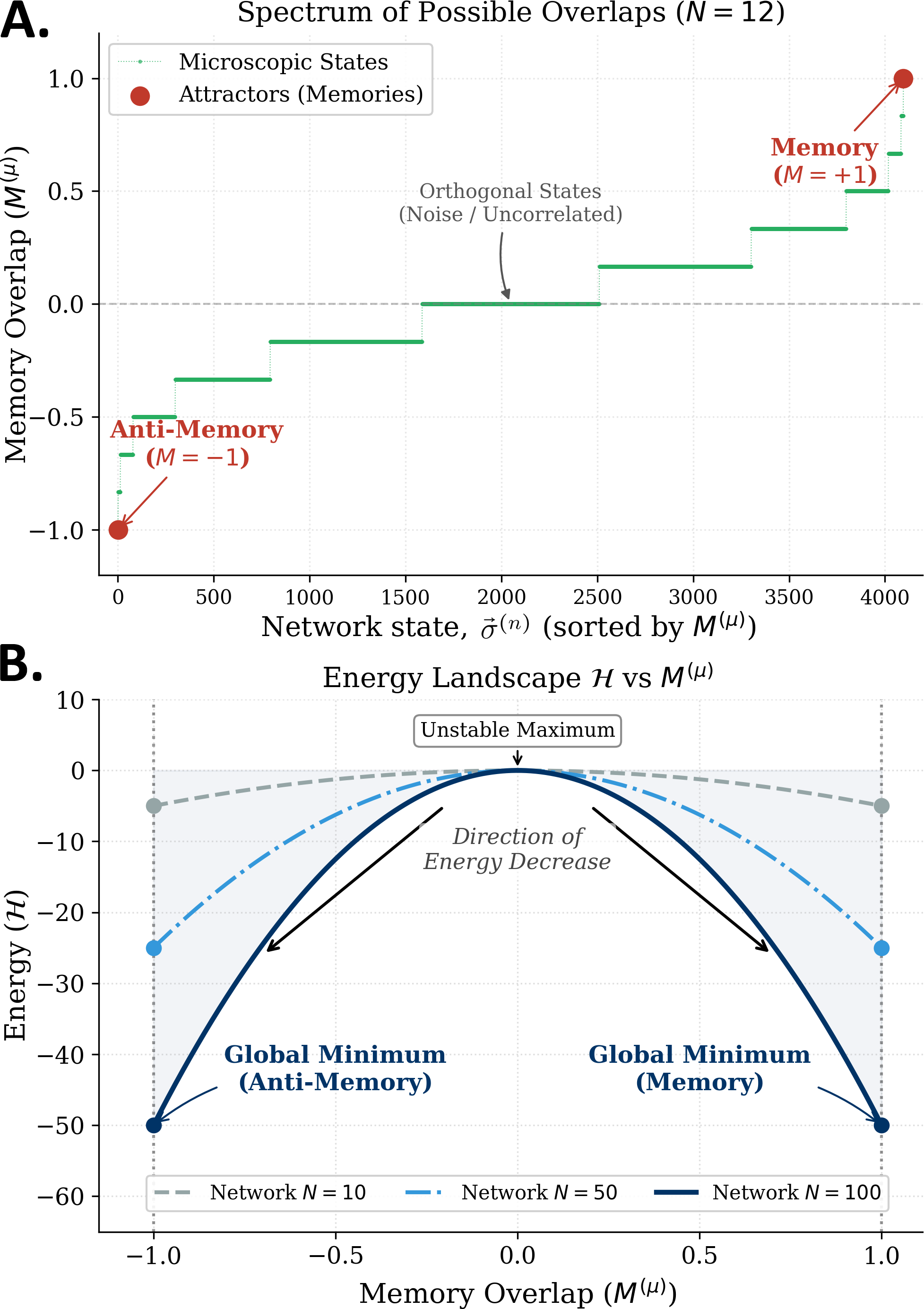}
    \caption{\label{sfig:overlapenergy}
    (a)~Overlap between all $2^{12}$ states and a single (random) memory pattern $\memstate$
    for a network with $N=12$ neurons, Eq.~\eqref{seq:overlap}.
    The microscopic states are sorted by their overlap values.
    The ``staircase'' structure reflects the discreteness of the system.
    Most of the states are uncorrelated with the memory, $\overlapmem\approx0$.
    The only states with $\overlapmem=\pm 1$ are the memory, $\netstate[n]=\memstate$, and the antimemory,
    $\netstate[n]=-\memstate$.
    (b) The inverted parabola, Eq.~\eqref{seq:energydef} for $P=1$
    is the simplest function that can be used to impose that $\overlapmem=\pm1$ are minima of $\Hfunc$.
    The arrows indicate the direction of energy decrease.}
\end{figure}

We define the normalized overlap between a network state $n$ and a memory $\mu$,
\begin{equation}
\label{seq:overlap}
\overlapmem[\memstate[\mu],\netstate[n]]=\dfrac{1}{N}\sum_{i=1}^N\spinmemstate[\mu]_i\spinnetstate[n]_i\ .
\end{equation}
Equation~\eqref{seq:overlap} is simply the normalized scalar product between $\netstate[n]$ and $\memstate[\mu]$ [Fig.~\ref{sfig:overlapenergy}(a)].
The normalization factor $1/N$ ensures that
$\overlapmem[\memstate[\mu],\netstate[n]]\in[-1,1]$.
When $\overlapmem[\memstate[\mu],\netstate[n]] = 1$,
the network state matches the memory exactly, $\netstate[n]=\memstate[\mu]$.
When $\overlapmem[\memstate[\mu],\netstate[n]] = -1$,
the network matches the corresponding antimemory, $\netstate[n]=-\memstate[\mu]$.
This antimemory arises from the spin-flip symmetry of the system, which makes only half of the $2^N$ states unique.

Then, we define the energy function~\cite{hertzRedesNeurais,peretto}
so that $M=\pm1$ can be minima [the case $P=1$ is shown in Fig.~\ref{sfig:overlapenergy}(b)]:
\begin{equation}
\label{seq:energydef}
\Hfunc = -\dfrac{1}{2}N \sum_{\mu=1}^P \left(\overlapmem[\memstate[\mu],\netstate]\right)^2\ .
\end{equation}
We use Eq.~\eqref{seq:overlap}
and expand the squared sum as
\begin{subequations}
\begin{align}
    \mathcal{H}(\vec{\sigma})
    &= -\frac{1}{2}N \sum_{\mu=1}^P \left( \frac{1}{N} \sum_{i=1}^N \spinmemstate[\mu]_i \spinnetstate_i \right)^2 \\
    &= -\frac{1}{2N} \sum_{\mu=1}^P \left[ \sum_{i=1}^N \sum_{j=1}^N (\spinmemstate[\mu]_i \spinnetstate_i)(\spinmemstate[\mu]_j \spinnetstate_j) \right]\\
    &= -\frac{1}{2} \sum_{i=1}^N\sum_{j=1}^N \left(\frac{1}{N} \sum_{\mu=1}^P \spinmemstate[\mu]_i \spinmemstate[\mu]_j \right) \sigma_i \sigma_j\ .
\end{align}
\end{subequations}
We define the interaction weight,
\begin{equation}
\label{seq:weighthopfield}
W_{ij} = \frac{1}{N} \sum_{\mu=1}^P \spinmemstate[\mu]_i \spinmemstate[\mu]_j\ ,
\end{equation}
and obtain the familiar energy function,
\begin{equation}
\label{eq:it}
\Hfunc[\netstate] = -\frac{1}{2} \sum_{i=1}^N\sum_{j=1}^N W_{ij} \sigma_i \sigma_j\ .
\end{equation}

We can further simplify Eq.~\eqref{eq:it}   by separating the main diagonal terms and noting that
$(\spinmemstate[\mu]_i)^2=(\spinnetstate[\mu]_i)^2=1$, so that $W_{ii}=P/N$
\begin{align}
\Hfunc[\netstate] &= -\frac{1}{2} \left( \sum_{i=j=1}^N \frac{P}{N}
                        + \sum_{i \neq j}^N W_{ij} \sigma_i \sigma_j \right)\\
                  &=-\frac{P}{2} -\frac{1}{2} \sum_{i \neq j}^N W_{ij} \sigma_i \sigma_j\ .
\label{seq:energyPconst}
\end{align}
The constant term in Eq.~\eqref{seq:energyPconst} gives a hint that we can make
$P\to P+\sum_i H_i\sigma_i$, introducing the local field $H_i$. This substitution
is not expected to change the dynamics as long as $\avg{\sum_i H_i \sigma_i}\sim$ constant.\cite{hertzRedesNeurais,peretto}
Stability of memories will impose a further restriction on $H_i$ (see main manuscript).

Also, note that the constant $P$ can be neglected because it does not change the dynamics.
Finally, we impose $W_{ii}=0$ to make $P$ naturally disappear from the energy function.
This assumption is not necessary, but it helps
us to mathematically guarantee convergence to a memory in the main manuscript.~\cite{hertzRedesNeurais}
After these modifications, we get
\begin{equation}
\label{seq:energygeneral}
\Hfunc[\netstate] = -\frac{1}{2} \sum_{i=1}^N \sum_{j=1}^N W_{ij} \sigma_i \sigma_j - \sum_{i=1}^N H_i \sigma_i\ .
\end{equation}
The local field $H_i$ is biologically interpreted as the activation threshold of neuron $i$.
If we wrote $J_{ij}=N\, W_{ij}$, the Hopfield Hamiltonian would be equal to that of the
spin glass, Eq.~\eqref{seq:SKHamilton}.

\section{Pedagogical applications across physics curricula}

The Hopfield model is particularly well suited for interdisciplinary
teaching because a single mathematical structure can be explored from multiple perspectives throughout the undergraduate physics curriculum.
Rather than introducing disconnected examples in different courses,
the same network can be revisited in progressively more sophisticated contexts,
allowing students to connect ideas from computation, dynamics, linear algebra, and statistical physics within a unified framework.

In the Supplementary Material, we provide a collection of example problems
that introduce the Hopfield model through hands-on exploration.
The activities are partially self-guided and encourage students
to investigate the network in a research-oriented way.
Topics include hallucinations (local minima of $\Hfunc$),
helping illustrate the mechanisms behind errors in AI systems,
including modern large language models.
We also provide example codes for training the network,
retrieving memories, and sorting network states to reveal
the attractor structure of the energy landscape.
A complete repository containing the simulation code
and worked examples is freely available.~\cite{Girardi2025HopfieldSim}

In computational physics, the Hopfield model provides an accessible introduction
to simulation-based reasoning.
Students can implement the dynamics from scratch using only basic programming tools,
while exploring concepts such as convergence, stability, metastability, and noise.
By starting from corrupted memories, they can study the robustness of information retrieval
and estimate the critical fraction of flipped neurons above which convergence fails.
The dynamics can be visualized through pattern evolution, overlap functions, and energy curves.
These simulations also reveal spurious memories and show how complex collective behavior
can emerge from simple deterministic update rules.

The same simulations naturally connect to nonlinear dynamics.
Memory states become attractors of the dynamics,
while their basins of attraction can be explored numerically
by evolving all possible initial conditions in small networks.
Students can therefore study high-dimensional attractor landscapes explicitly,
going beyond traditional low-dimensional phase portraits.
Visualizations of the Hamiltonian organized by attractor basin reveal
how stable memories correspond to energy minima,
while unstable retrieval and hallucinations emerge from competing local minima.
Concepts such as convergence, stability, and basins of attraction
thus become directly linked to the geometry of the energy landscape.

The Hopfield model also provides concrete applications of linear algebra.
Memory states are represented as vectors,
overlaps correspond to scalar products,
and the Hebbian learning rule constructs the connectivity matrix
as a sum of outer products.
Instead of treating matrices and vector spaces as purely abstract objects,
students see them operating directly in a problem of information storage and retrieval.
For example, the stability of memories can be interpreted geometrically
through the near-orthogonality of memory vectors,
connecting linear algebraic ideas to dynamical behavior.

Finally, the Hopfield network provides a natural bridge to statistical physics.
Its Hamiltonian has the same structure as spin models,
allowing concepts such as energy landscapes,
thermal fluctuations, and phase transitions
to be introduced in a familiar setting while connecting them
to neuroscience and artificial intelligence.
Our code can also be modified to include temperature and noise.
Students can then observe how thermal fluctuations destabilize memory retrieval,
driving the network from ordered memory states to disordered configurations,
analogous to the paramagnetic phase in magnetic systems.
Likewise, storing progressively more memories modifies the energy landscape
in a way reminiscent of frustration and competing minima in spin glasses.

Because all these topics emerge from the same simple model,
the Hopfield network can serve as a unifying pedagogical platform throughout the physics curriculum.
The same code developed in an introductory computational project can later be reused
to discuss attractors in dynamical systems,
matrix representations in linear algebra,
and phase transitions in statistical mechanics.
This continuity helps students connect concepts that are often taught as separate subjects.

\newpage
\section{Simple python code for storage and retrieval of memory}
\label{app:code}

We include a \verb|Python 3.8.2| implementation of the Hopfield model for Hebbian-like storage and retrieval algorithms.
The implementation depends only on \verb|numpy|, a standard library for numerical calculations in \verb|python|.
It was tested with \verb|numpy 1.24.4|.
More details on the code repository~\cite{Girardi2025HopfieldSim}
in \url{https://github.com/neuro-physics/hopfield-neural-network}.
To make the code clearer in this paper, we changed the function names from the ones originally given in the repository.
\vfill
\par\newpage\clearpage
\onecolumngrid
\textbullet\ Storage of memory, also known as \textit{training the network}:

\begin{lstlisting}[language=Python]
import numpy as np
def store_patterns(patterns):
    """
    This function calculates the interaction matrix Wij
    from a given list of patterns to be stored,
    using standard outer-product rule.
    Parameters
    ----------
    patterns : array-like or list of numpy.ndarray
    A collection of reference patterns used to train the Hopfield network.
    Each pattern is  a 1D array of length N (flattened vector). 
    Values are +1 or -1.
    Returns
    -------
    W : numpy.ndarray
    The weight matrix of shape (N, N), initialized according to Hebbian learning.
    """
    patterns = np.atleast_2d(patterns)
    N = len(patterns[0])
    W = np.zeros((N, N))
    for xi in patterns:
        W += np.outer(xi, xi) / N
    np.fill_diagonal(W, 0)
    return W
\end{lstlisting}

\par\newpage  
\textbullet\ Hopfield dynamics for memory retrieval given an initial condition, also known as \textit{pattern matching}:
\begin{lstlisting}[language=Python]
import numpy as np
def get_memory_async(W, s0, max_MCsteps):
    """
    This function iterates the Hopfield dynamics (zero-temperature)
    from a given initial configuration s0, returning the
    state that the network converged to.
    Parameters
    ----------
    W : numpy.ndarray
        Symmetric weight matrix of shape (N, N), where N is the number of neurons.
    s0 : numpy.ndarray
        Initial state vector of shape (N,), with entries +1 or -1.
    max_MCsteps : int, optional (default=10)
        Maximum number of Monte Carlo steps (epochs). Each step updates all neurons once.
    Returns
    -------
    s : numpy.ndarray
        Final state vector after convergence or reaching max_MCsteps.
    """
    # the number of neurons
    N = len(s0) 
    # vector containing the index of each neuron
    indices = np.arange(N)
    # copy of the initial state vector
    s = s0.copy().astype(float)
    # main time loop
    # t_MC = 1 MC step = 1 epoch = N time steps
    for t_MC in range(1,max_MCsteps):
        # Randomize neuron update order
        np.random.shuffle(indices)
        # flag to keep track of the network state
        state_changed = False
        # try to update every neuron
        for i in indices:
        # Calculate the local field for neuron i
        h_i     = np.dot(W[i, :], s)
        # new state of neuron i
        s_i_new = 1.0 if h_i >= 0 else -1.0
        # if the state of neuron i changed
        if s_i_new != s[i]:
            # keep the new state
            s[i]  = s_i_new
            # update network state flag
            state_changed = True
        # no neurons changed state during a full pass,
        # we hit a local minimum
        # so we exit and return the retrieved pattern
        if not state_changed:
            break
    return s
\end{lstlisting}

\par\newpage  
\textbullet\ Sorting states in basins of attraction:
\begin{lstlisting}[language=Python]
import numpy as np
def sort_basins(state_mem, E):
    """
    Group states by attractor basin and sort the states inside each basin
    according to their energy.
    Parameters
    ----------
    state_mem : ndarray of shape (n_states,)
        Integer array whose element ``state_mem[i]`` gives the index mu 
        of the attractor (memory) reached by state ``i`` under the network dynamics.
    E : ndarray of shape (n_states,)
        Energy associated with each network state.
    Returns
    -------
    states_basin : ndarray of shape (n_states,)
        Array containing the reordered state indices. States are first
        grouped by attractor basin and then sorted by increasing energy
        within each basin.
    E_basin : ndarray of shape (n_states,)
        Energies reordered consistently with ``states_basin``.
    Notes
    -----
    The ordering produced by this function is useful for visualizing the
    energy landscape of Hopfield networks, since states belonging to the
    same attractor basin appear grouped together.
    """
    # first group states by attractor index
    states_basin = np.argsort(state_mem)
    # reordered energies
    E_basin = E[states_basin]
    # unique basin labels
    basins = np.unique(state_mem)
    # sort states inside each basin by energy
    for mu in basins:
        # positions inside the reordered array
        ind = np.where(state_mem[states_basin] == mu)[0]
        # energy ordering inside this basin
        k = np.argsort(E_basin[ind])
        # reorder basin states
        states_basin[ind] = states_basin[ind][k]
        E_basin[ind]      = E_basin[ind][k]
    return states_basin, E_basin
\end{lstlisting}

\par\newpage  

\end{document}